\documentclass{aa}
\usepackage{longtable}
\usepackage{graphicx}
\usepackage{hyperref}
\usepackage{txfonts}
\newcommand{\placefigone}{
\begin{figure*}
    \includegraphics[width=\hsize]{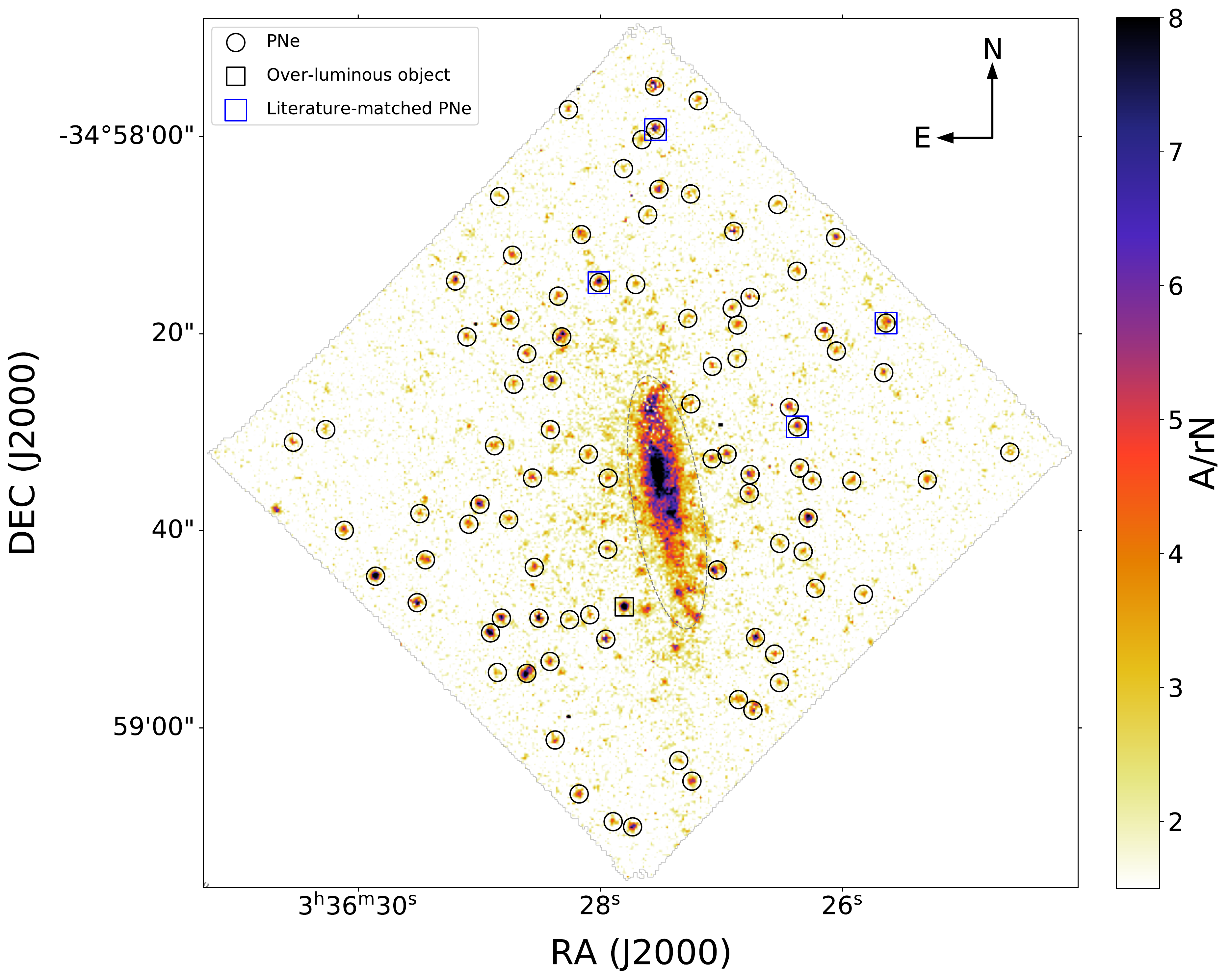}
    \caption{FCC\,167: Map of the peak amplitude to residual-noise level ratio (A/rN) of the [\ion{O}{iii}]5007 line, based on our spaxel-by-spaxel fit for the [\ion{O}{iii}] doublet in the emission datacube. The sources detected and labelled PNe are shown by a black circle. The over-luminous object (see Sect. 5.1) is highlighted by a black square. The PNe that mach those reported by \citet{Feldmeier2007CalibratingResults} are highlighted by blue squares.} The dashed ellipsoid marks the central region that was disregarded owing to the presence of diffuse ionised-gas emission \citep[see also][]{Viaene2019The167}.
    \label{fig:FCC167_A_rN}
\end{figure*}
}

\newcommand{\placefigtwo}{
\begin{figure*}
    \includegraphics[width=\hsize]{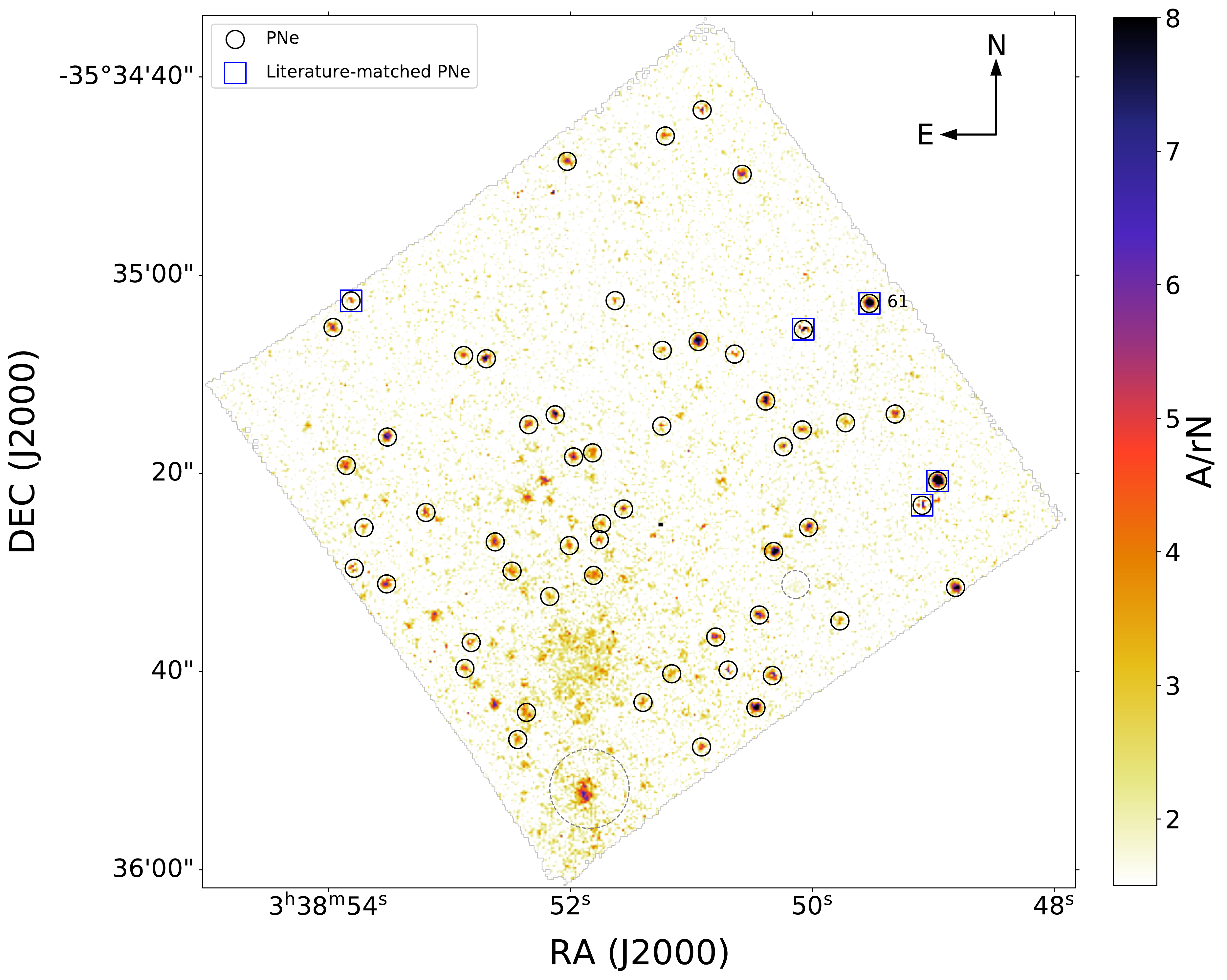}
    \caption{FCC\,219: Similar to Fig.\ref{fig:FCC167_A_rN}, but show the PNe within FCC\,219. Sources are highlighted by black circles,and blue square indicate the sources that we matched with those of \citet{McMillan1993PlanetaryCluster} within the FOV. The central mask, located towards the south of the FOV (dashed circle), excludes regions affected by diffuse ionised-gas emission \citep[see][]{Iodice2019TheMaps}. A foreground star is also masked out. This is indicated by the small dashed circle to the right of the FOV.}
    \label{fig:FCC219_A_rN}
\end{figure*}
}

\newcommand{\placefigthree}{
\begin{figure*}
    \centering
    \includegraphics[width=\hsize]{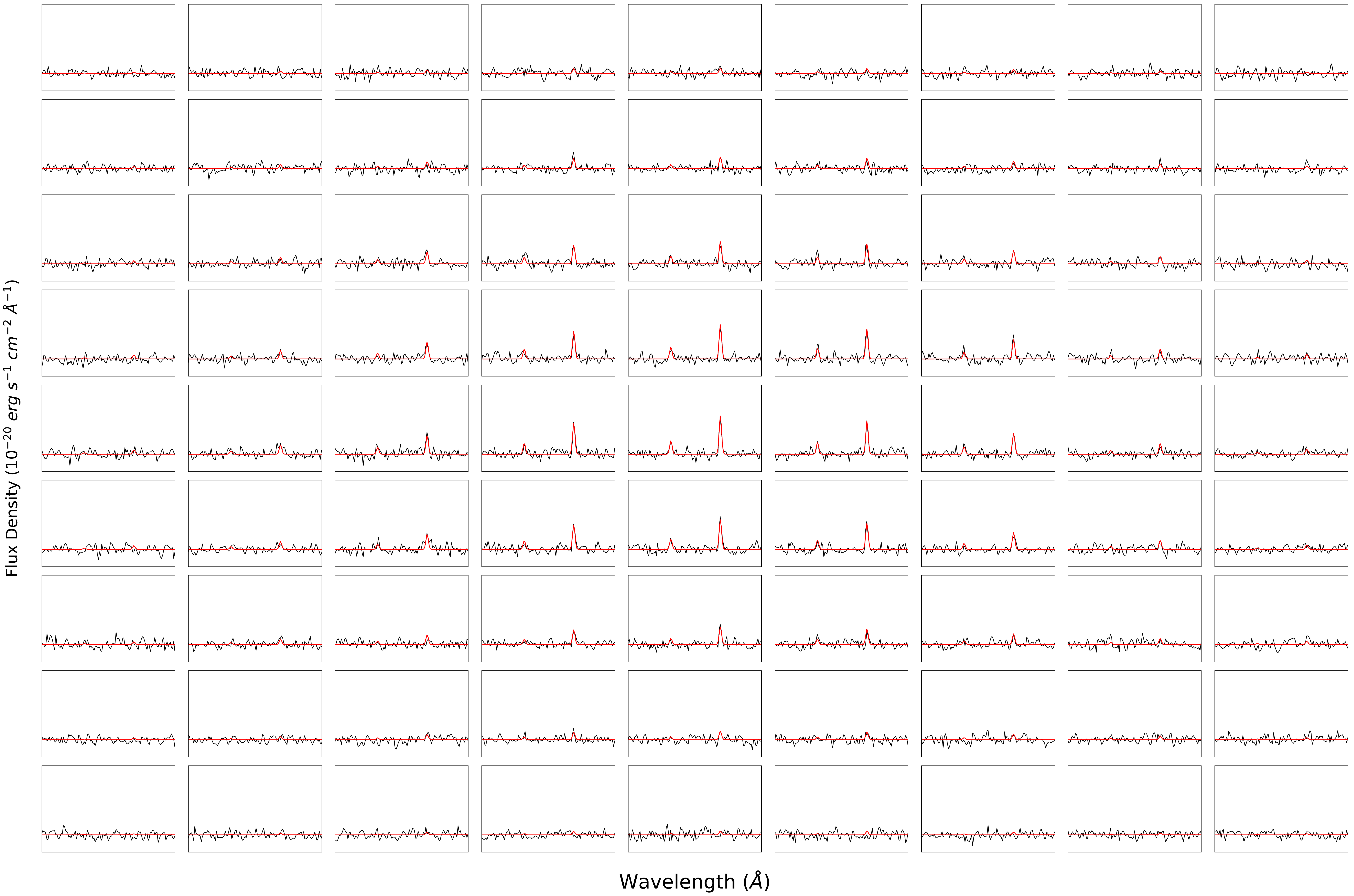}
    \caption{Example of an outcome from our 3D fitting for one PNe source (61), located in the central region of FCC\,219 (see Fig. \ref{fig:FCC219_A_rN}). For each of the $9 \times\ 9$ spaxels plotted, the corresponding wavelength range spans 4950-5080 \AA. The scale of the y-axis is chosen arbitrarily to best illustrate our fits. Spectral data are shown in black, and our [\ion{O}{iii}] model is shown in red. The entire 9x9 spaxel region is displayed to highlight the expected variation in signal (central pixels) and noise (outer pixels). Each spaxel corresponds to a spatial scale of 0.2 arcsecond.}
    \label{fig:FCC219_spaxel_by_spaxel}
\end{figure*}
}

\newcommand{\placefigfour}{
\begin{figure}
    \includegraphics[width=\hsize]{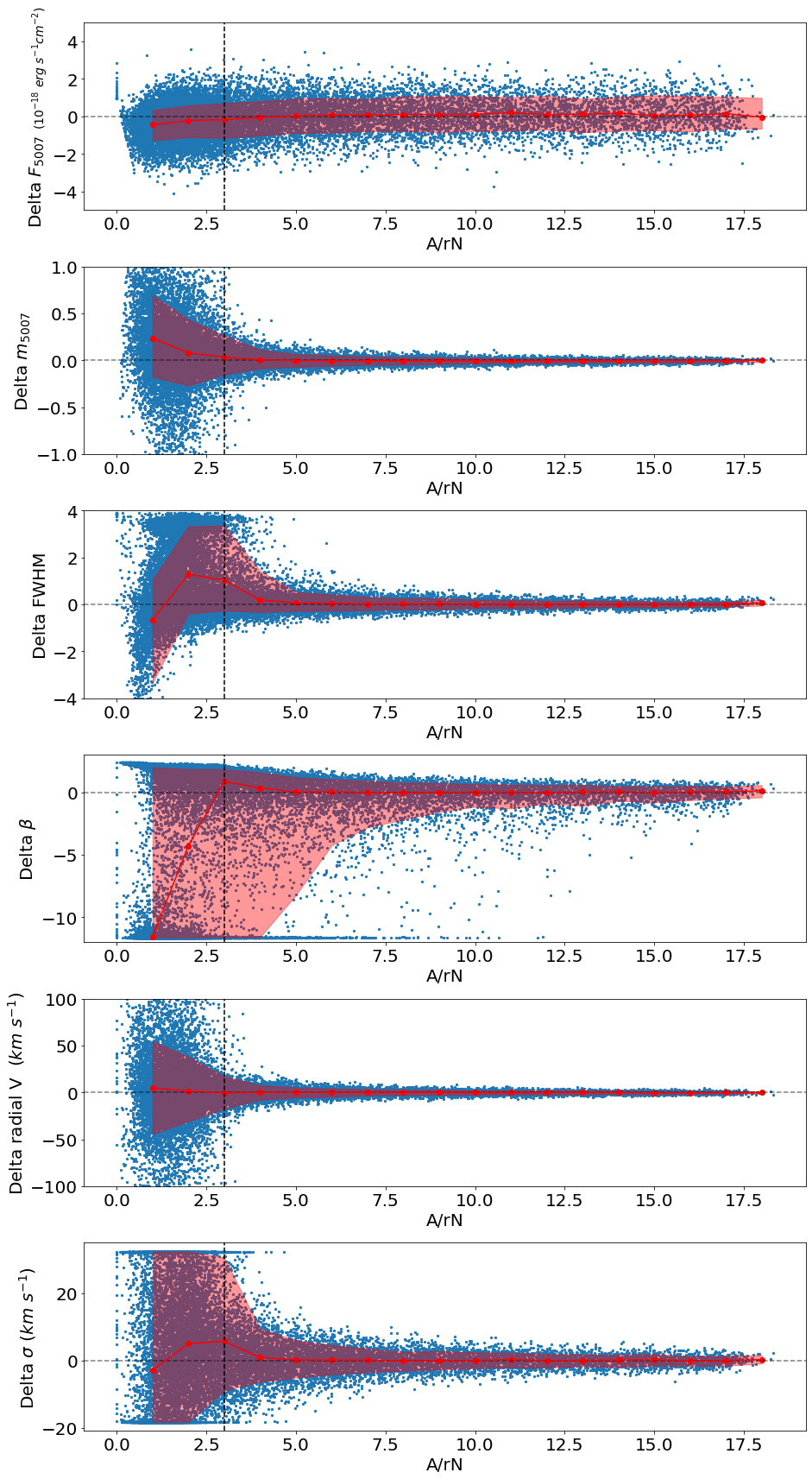}
    \caption{Simulation of the detection and retrieval of PSF parameters FWHM and $\beta$, and determination of the accuracy of the model fits for the total flux of a source. The blue points show the individual simulation results, and the red points show the median values, binned in A/rN. The upper and lower parts of the red region indicate the 86th and 16th percentile, respectively. Top row, first panel: delta [\ion{O}{iii}] flux. Top row, second panel: delta $\rm M_{5007}$. Top row, third panel: delta FWHM. Top row, fourth panel: delta $\beta$. Top row, fifth panel: delta radial velocity derived from wavelength position. Top row, sixth panel: delta velocity dispersion of the [\ion{O}{iii}] emission lines.}
    \label{fig:free_PSF_sims}
\end{figure}
}

\newcommand{\placefigfive}{
\begin{figure}
    \includegraphics[width=\hsize]{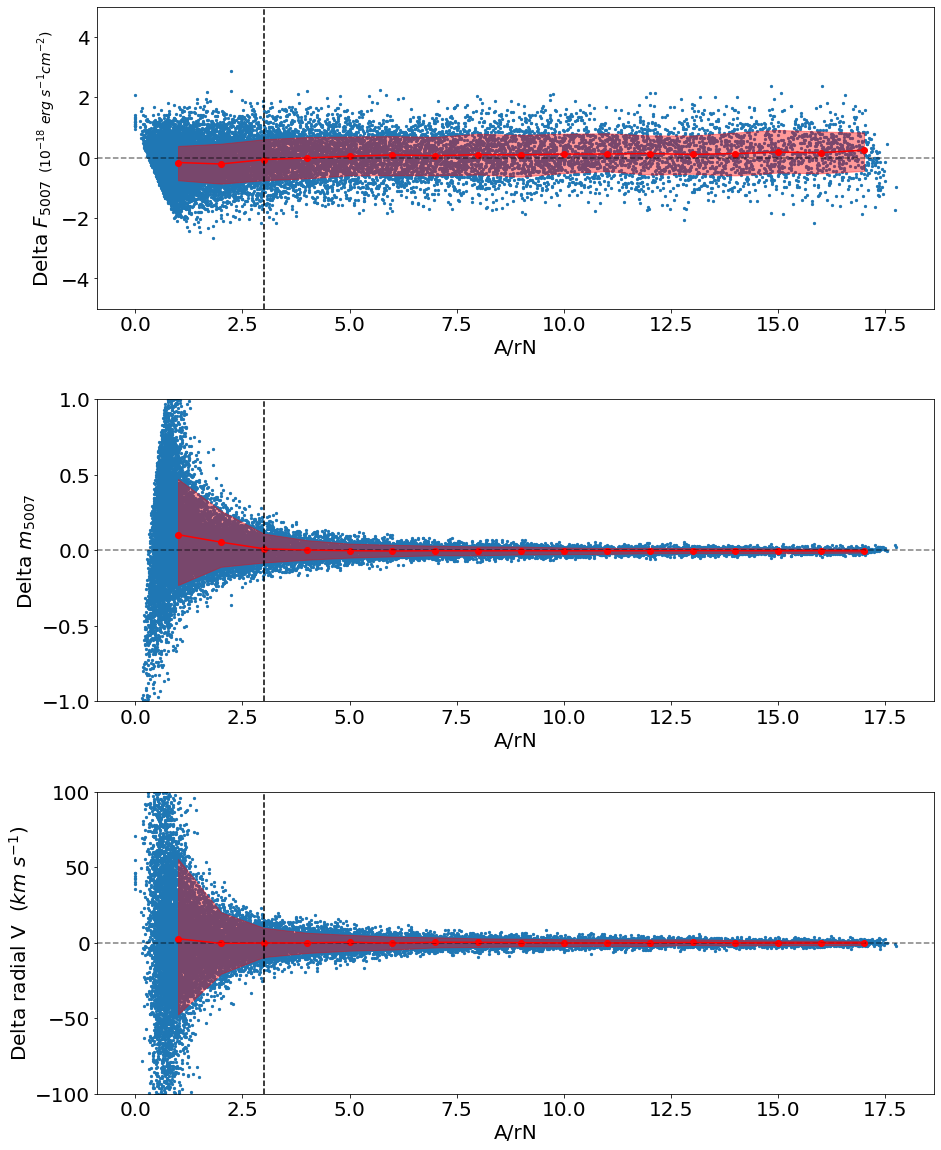}
    \caption{Simulation results when the PSF values of FWHM and $\beta$ are known and held constant. Top row, first panel: Delta Flux of [\ion{O}{iii}] against fitted A/rN value of each source. Top row, second panel: Delta $\rm M_{5007}$ against source A/rN. Top row, third panel: Delta radial velocity as measured from the offset of the [\ion{O}{iii}] emission line. The red points are the median value binned by A/rN. The upper and lower parts of the red region indicate the 86th and 16th percentile, respectively.}
    \label{fig:fixed_PSF_sims}
\end{figure}
}

\newcommand{\placefigsix}{
\begin{figure}
    \includegraphics[width=\hsize]{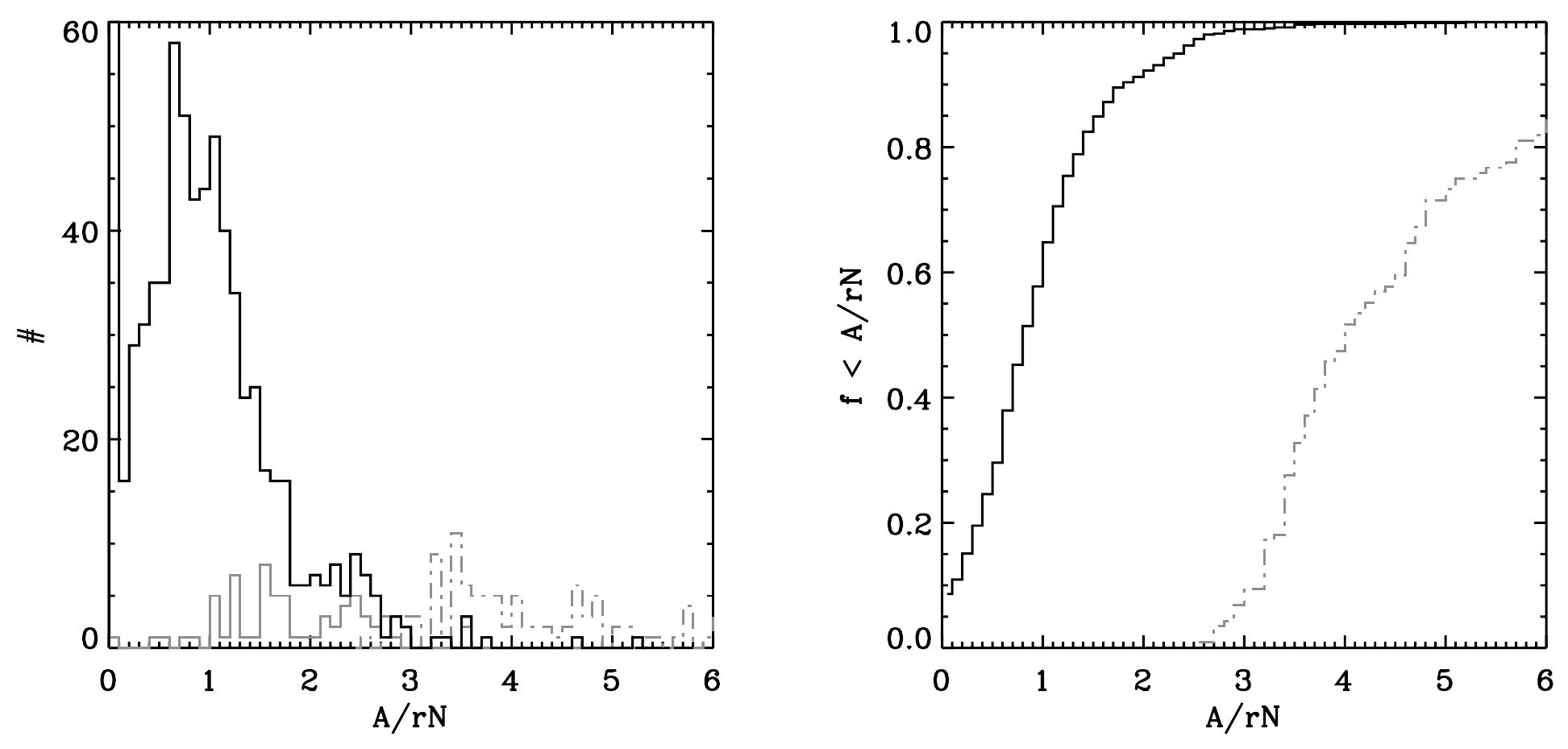}
    \caption{Central A/rN values from 1D+2D-fits to regions devoid of emission in FCC\,167. Left: Observed distribution of false-positive A/rN values. The grey line shows the values corresponding to regions closer to the centre, where template mismatches in the emission line cube systematically bias the A/rN values to higher values. For comparison, the dot-dashed grey lines also show the distribution of A/rN values for our candidate sources. Right: Cumulative distribution for the false-positive A/rN values, 99\% of which lie below A/rN=3. A/rN values corresponding to poor fits are excluded.} \label{fig:FCC167_falsepositive}
\end{figure}
}

\newcommand{\placefigseven}{
\begin{figure}
    \includegraphics[width=\hsize]{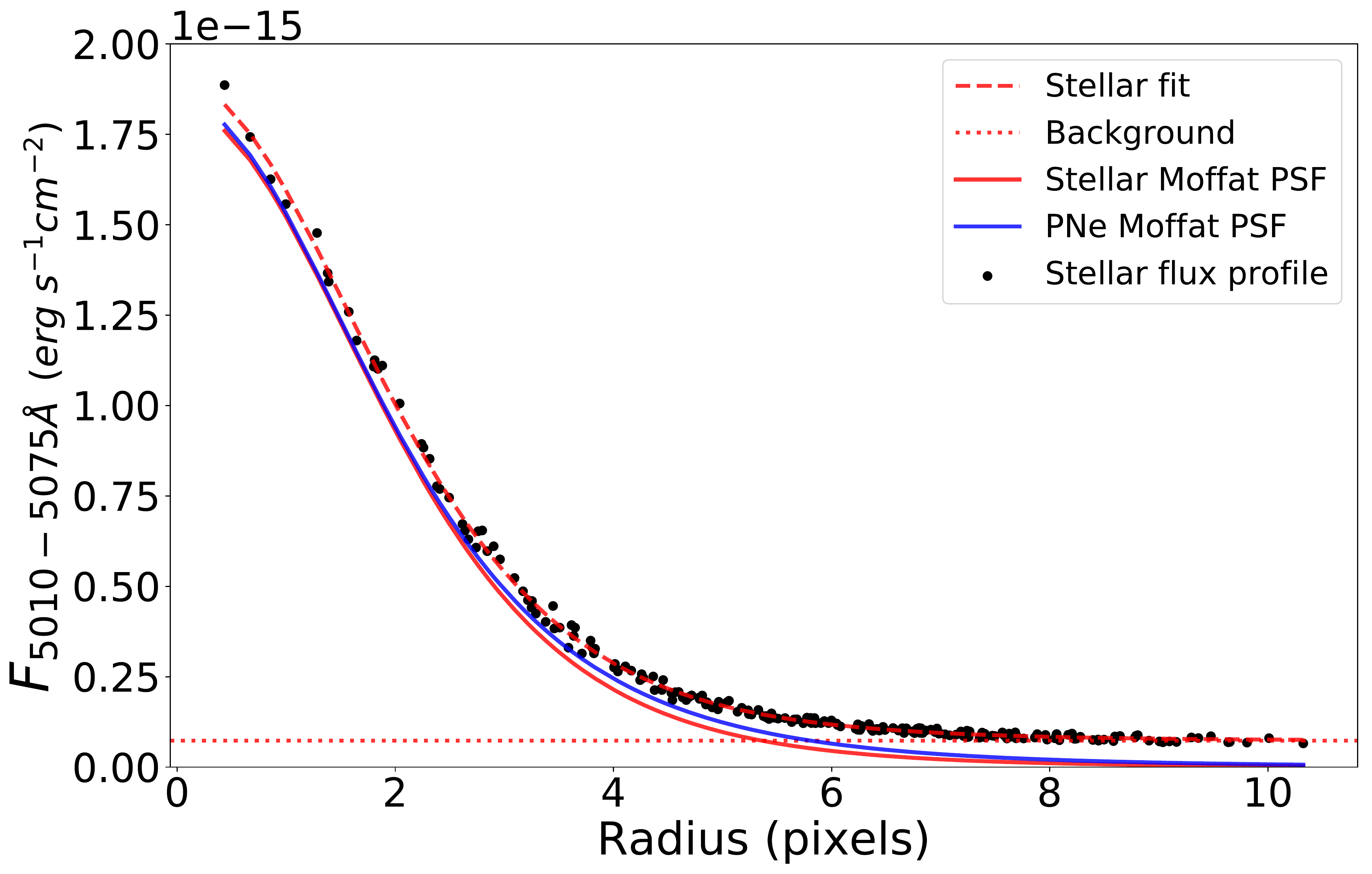}
    \caption{Radial profile of the star within the FOV of FCC\,219 (black dots). The best-fit model to the stellar light is shown by the red dashed line, and the red dotted line indicates the background level of galaxy light. The solid red and blue lines depict the actual PSF, without background, as described by the star and PNe, respectively. For the total flux and profile, this comparison highlights that the two approaches agree well.}
    \label{fig:PSF_star_vs_PNe}
\end{figure}
}

\newcommand{\placefigeightA}{
\begin{figure*}
    \includegraphics[width=\hsize]{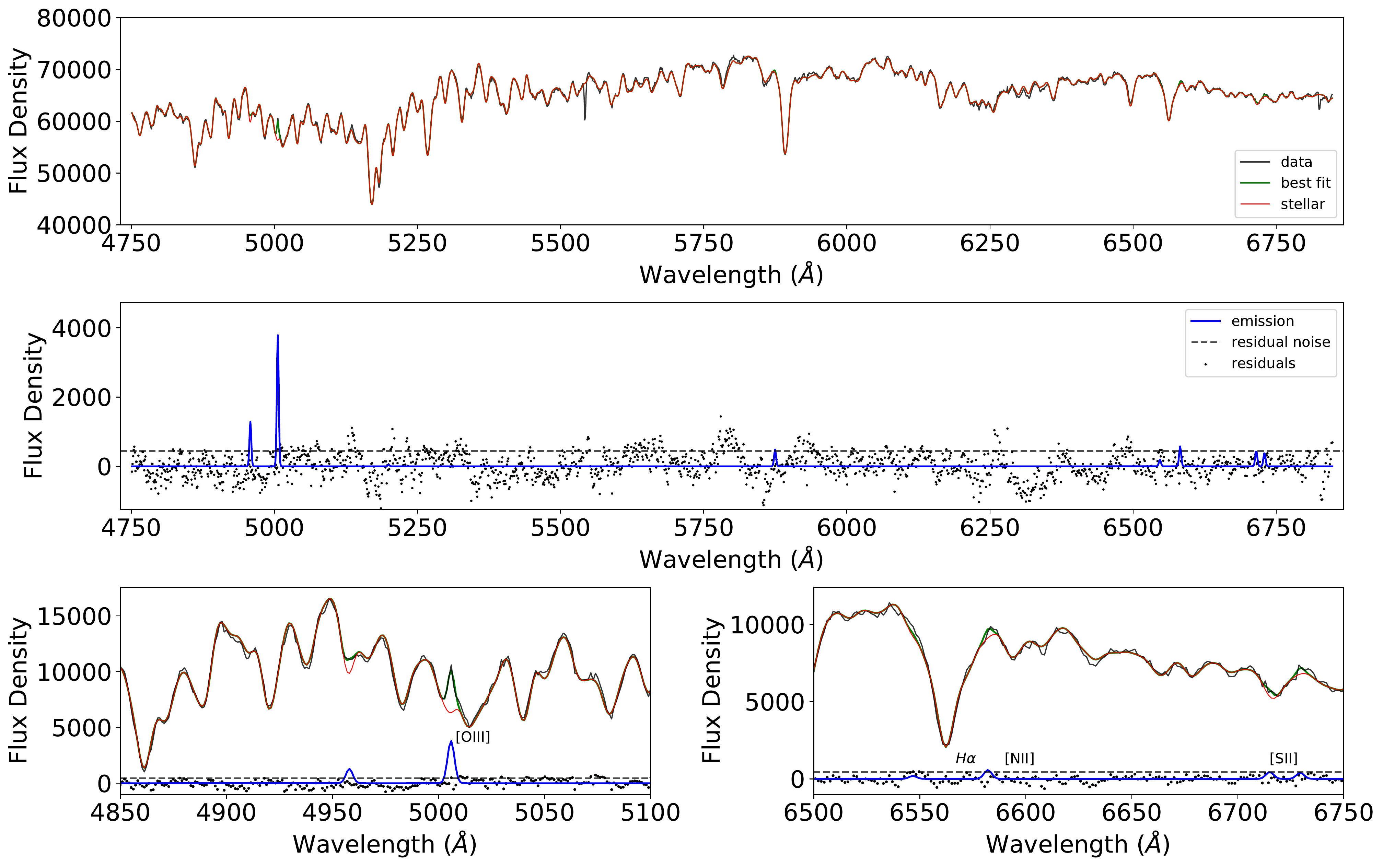}
    \caption{GandALF fit to the typical spectrum (green) of a PNe source (black line, top plot) from FCC\,167 (F3D J033627.08-345832.69), and its galaxy stellar background (red), showing strong [\ion{O}{iii}] lines and some $\rm H\alpha$ emission. The middle plot shows the emission lines as detected by GandALF (blue). The dashed horizontal line indicates the level of residual noise (standard deviation of the residuals from stellar subtraction (black points)). The lower left panel zooms into the H$\beta$ and [\ion{O}{iii}] doublet wavelength region, and the lower right panel shows the region occupied by $\rm H\alpha$ and the [\ion{N}{ii}] and [\ion{S}{ii}] doublets. The data, best fit, and stellar spectra shown in the bottom two plots are subtracted by an arbitrary number to better present and compare the fit of the nebulous and stellar emissions within each region.}
    \label{fig:gandalf_PNe}
\end{figure*}
}

\newcommand{\placefigeightB}{
\begin{figure*}
    \includegraphics[width=\hsize]{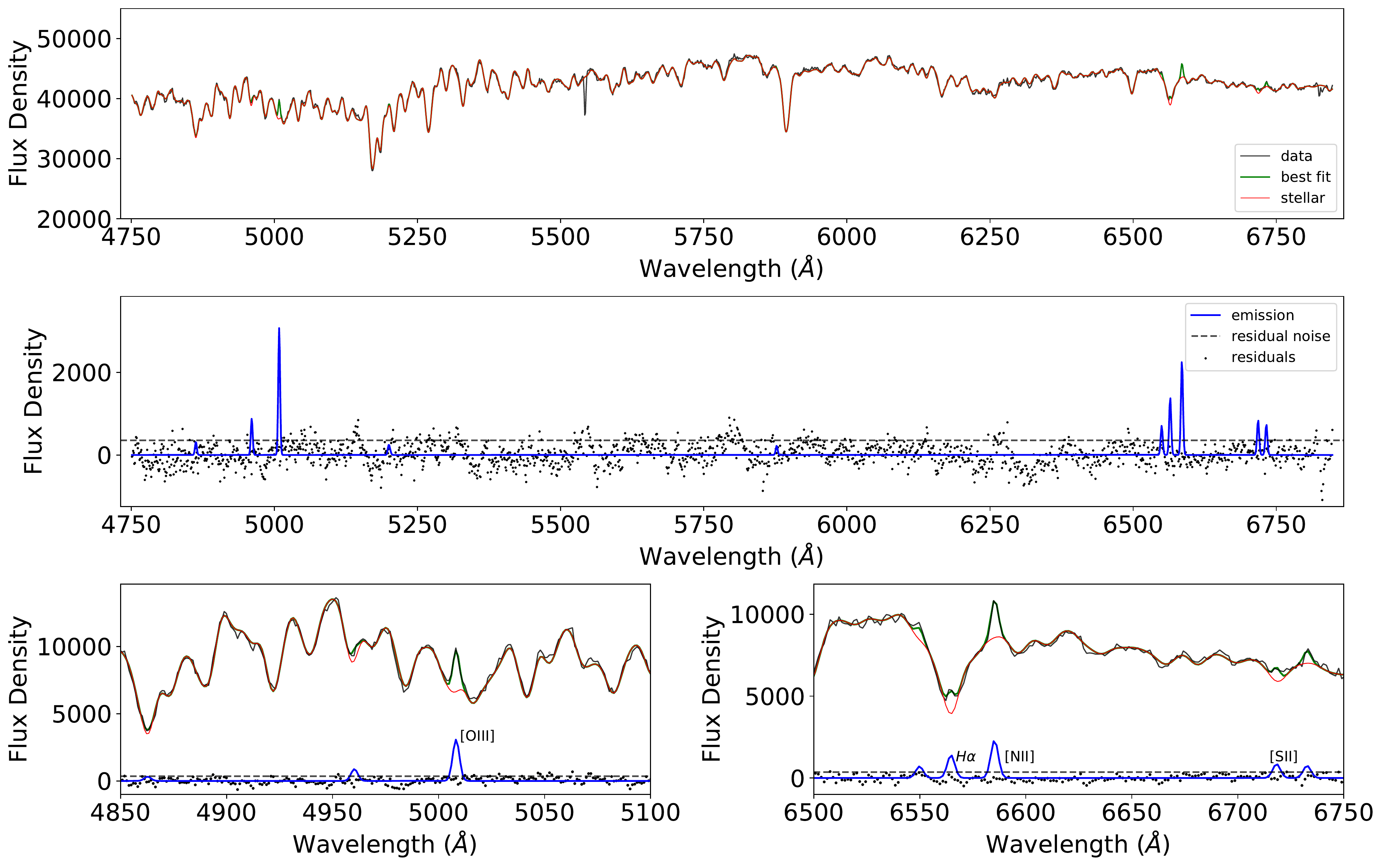}
    \caption{Same as Fig~\ref{fig:gandalf_PNe}, for a potential supernova remnant source within FCC\,167 (F3D J033627.66-345844.20). The detected $\rm H\alpha$, [\ion{N}{ii}] and [\ion{S}{ii}] emission is similar to that of [\ion{O}{iii}].}
    \label{fig:gandalf_SNR}
\end{figure*}
}

\newcommand{\placefiglita}{
\begin{figure}
    \includegraphics[width=\hsize]{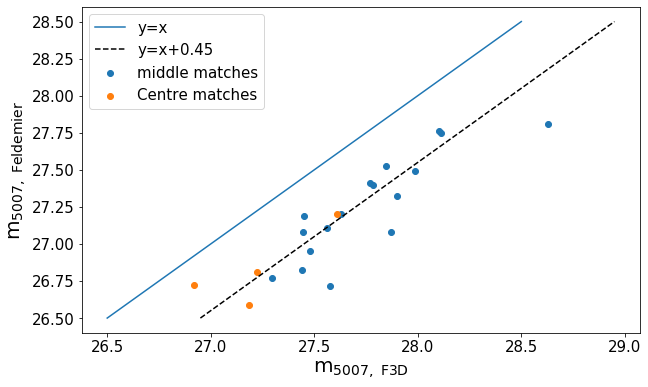}
    \caption{Comparison of the $m_{5007}$ of PNe detected within FCC\,167 (x-axis), with those of \citet{Feldmeier2007CalibratingResults} sample (y-axis). We match 21 sources that were found in the central (orange) and middle (blue) pointings of the F3D FCC\,167 observations. We find the comparisons to be consistent, but there is a systematic offset (dashed black line) in values, where our measured $m_{5007}$ values are $\sim$ 0.45 mag fainter than those of \citet{Feldmeier2007CalibratingResults}.}    \label{fig:FCC167lit}
\end{figure}
}

\newcommand{\placefiglitb}{
\begin{figure}
    \includegraphics[width=\hsize]{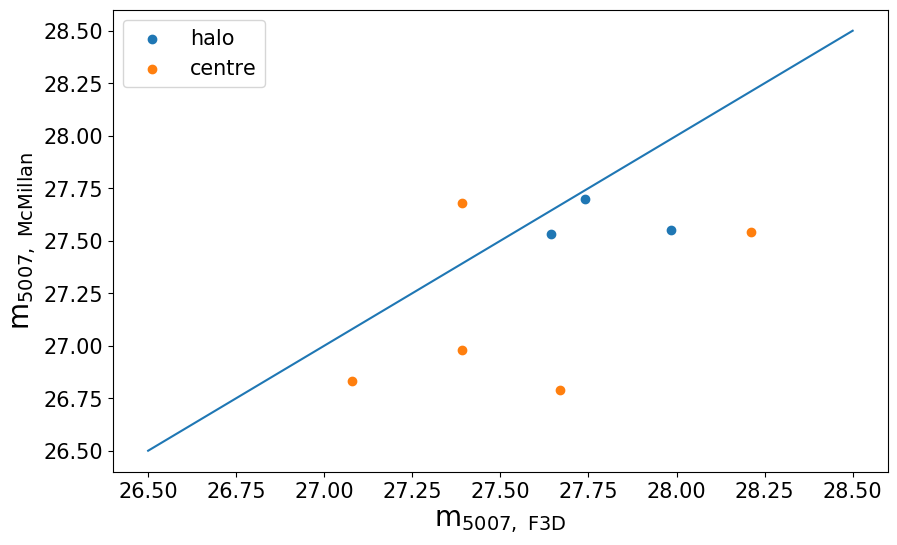}
    \caption{Comparison of PNe detected in FCC\,219 as detected here that match those reported by \citet{McMillan1993PlanetaryCluster}.}
    \label{fig:FCC219lit}
\end{figure}
}

\newcommand{\placefigvela}{
\begin{figure}
    \includegraphics[width=\hsize]{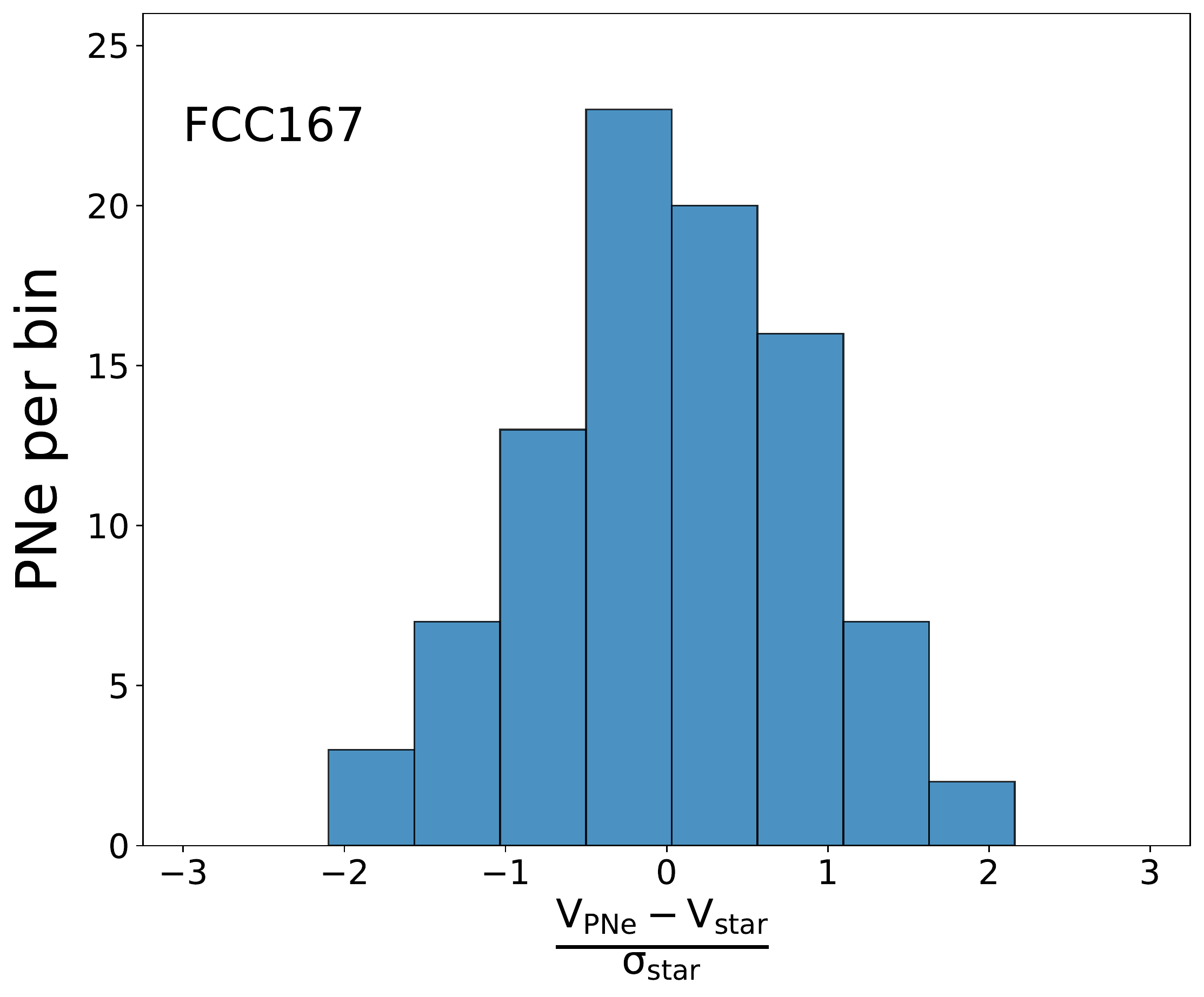}
    \caption{Values and distribution for the $\Delta V/\sigma$ ratio for PNe within FCC\,167, where $dV$ is the difference between the PNe velocities and that of the stars in the galaxy at the PNe location, and $\sigma$ is the stellar velocity dispersion, also at the PNe position.}
    \label{fig:FCC167_vel_hist}
\end{figure}
}

\newcommand{\placefigvelb}{
\begin{figure}
    \includegraphics[width=\hsize]{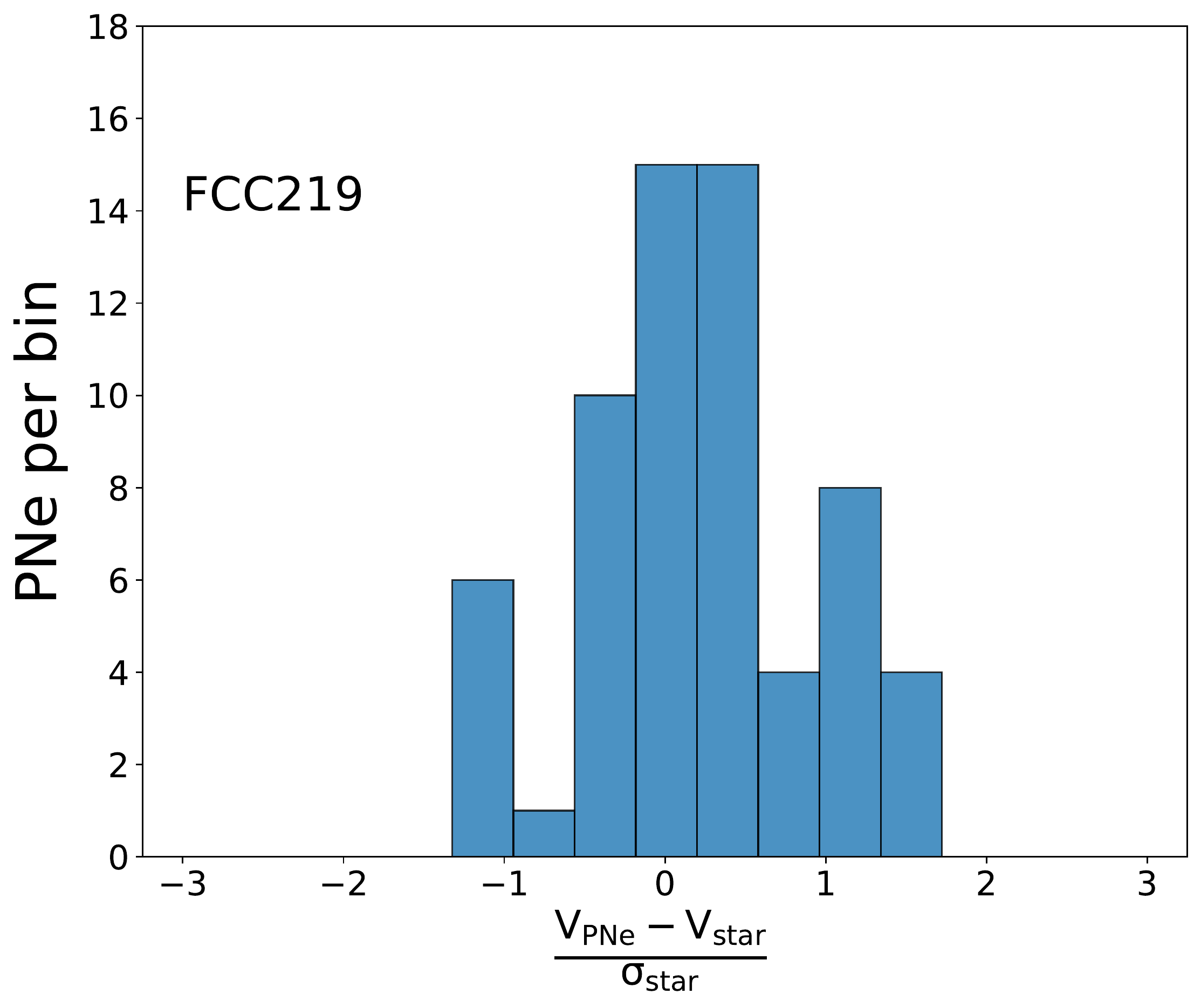}
    \caption{Distribution of $\Delta V/\sigma$ in FCC\,219. Similar to Fig.~\ref{fig:FCC167_vel_hist}.}
    \label{fig:FCC219_vel_hist}
\end{figure}
}

\newcommand{\placefigcontaminationa}{
\begin{figure}
    \includegraphics[width=\hsize]{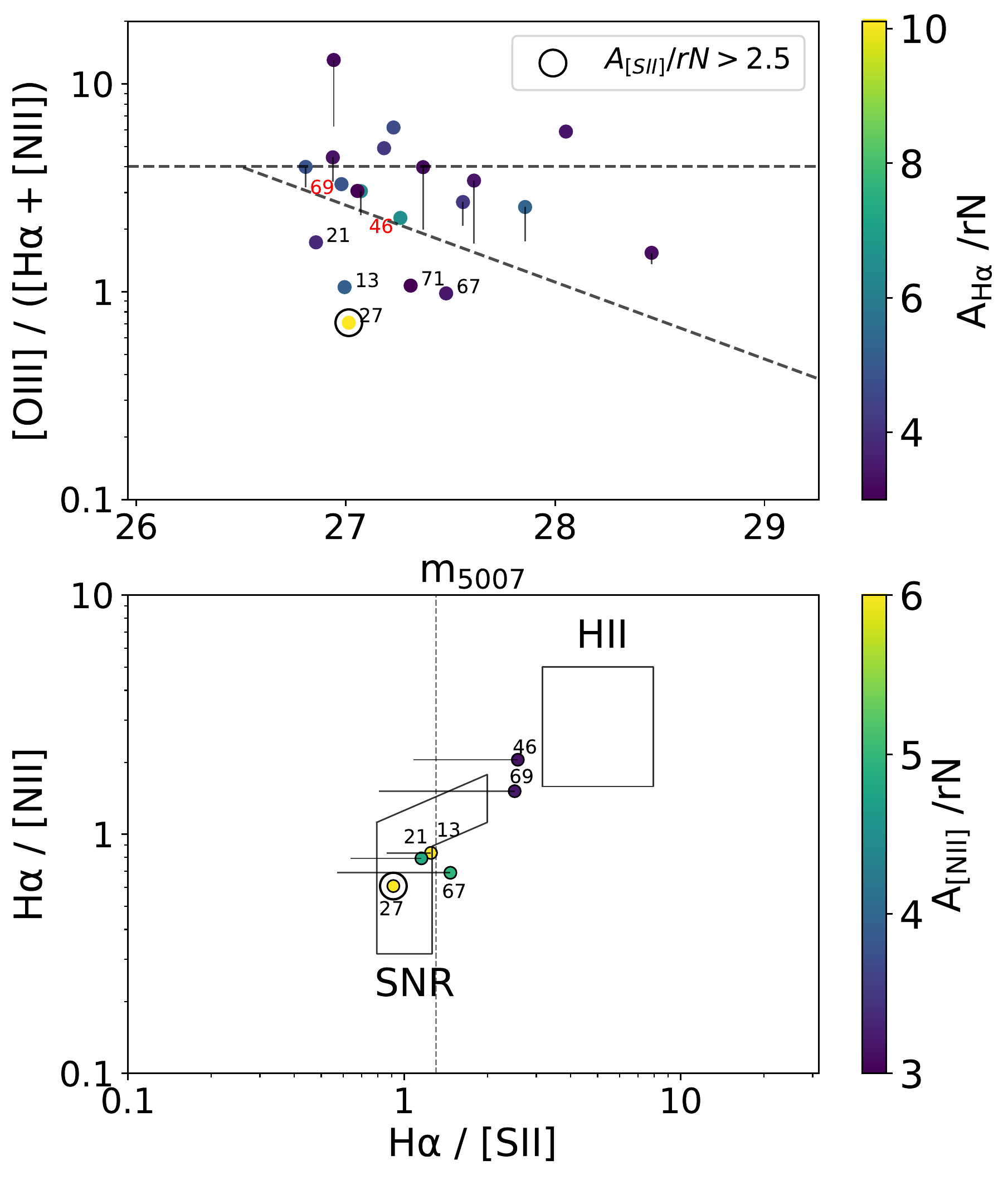}
    \caption{Contamination diagrams for FCC\,167. {\it Top panel:} Values for the [\ion{O}{iii}]/($\rm H\alpha$+[\ion{N}{ii}]) line ratio for the PNe candidate sources, as returned from our spectral fit and only for objects that were already validated for fit quality and detectability. The symbols are colour-coded according to the level of the signal-to-noise ratio of H$\alpha$, which is the predominant line of the H$\alpha$ [\ion{N}{ii}] pairing. The sources included here passed either the $A_{H\alpha}/rN$>3 or the $A_{[\ion{N}{ii}]}/rN$>3 filter. When this is not the case, a vertical line points to the corresponding lower limit for [\ion{O}{iii}]/($\rm H\alpha$+[\ion{N}{ii}]), assuming an upper limit in the [\ion{N}{ii}] flux corresponding to a $A_{[\ion{N}{ii}]+H\alpha}/rN = 3$. When a distance modulus of 31.24 mag is assumed (Sect.~5), the dashed lines show the region typically occupied by PNe according to \citet{Ciardullo2002PlanetaryScales} and \citet{Herrmann2008PlanetaryDistances}. {\it Lower panel:} Position of the PNe sources with firmly detected [\ion{N}{ii}] and H$\alpha$ emission in the \citet{Sabbadin1977SharplessNebula.} diagnostic diagram in which the regions occupied by PNe (from \citet{Riesgo2006RevisedNebulae}, SNRs and unresolved \ion{H}{ii}-regions. Similar to the top panel, horizontal lines indicate the range of values down to a lower limit for the $\rm H\alpha$/[\ion{S}{ii}] ratio where the [\ion{S}{ii}] doublet is not formally detected. Sources detected with \ion{S}{ii}/rN >3 are highlighted by a circle. One such source is found within FCC\,167. Sources are numbered to show where they lie in relation to each other within the two diagnostic diagrams.}
    \label{fig:PNe_impostor167}
\end{figure}
}

\newcommand{\placefigcontaminationb}{
\begin{figure}
    \includegraphics[width=\hsize]{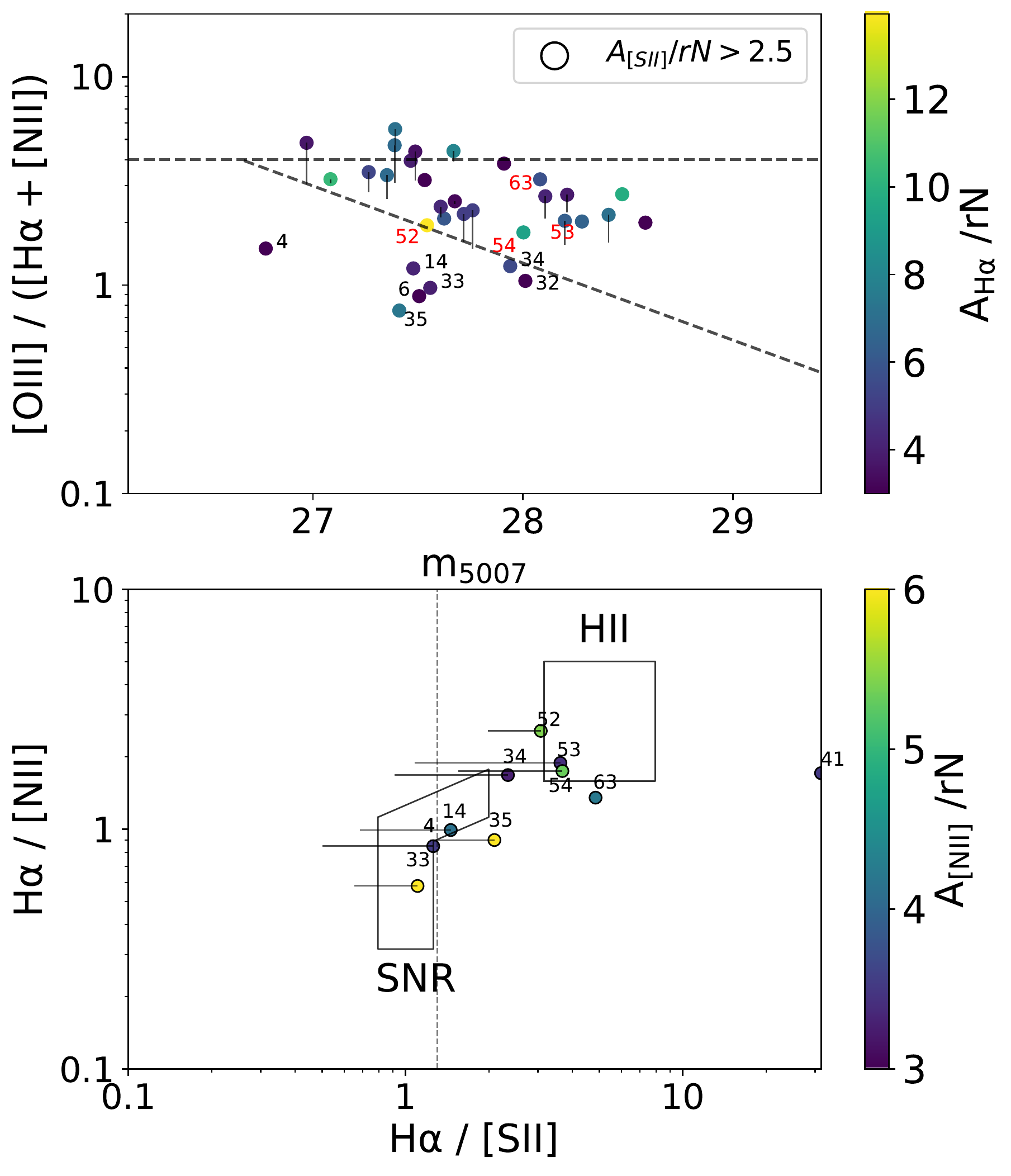}
    \caption{Contamination diagrams for FCC\,219 obtained with the same procedure and diagnostics as in Fig.~\ref{fig:PNe_impostor167}. The top panel shows that fewer sources reside outside of the dashed lines (from Eq.~\ref{eq:contamination}), and no sources exhibit [\ion{S}{ii}] emission at a signal-to-noise higher than 4. The points are again numbered to help identify sources in the two plots.}
    \label{fig:PNe_impostor219}
\end{figure}
}

\newcommand{\placefigPNLFa}{
    \begin{figure}
    \centering
    \includegraphics[width=\hsize]{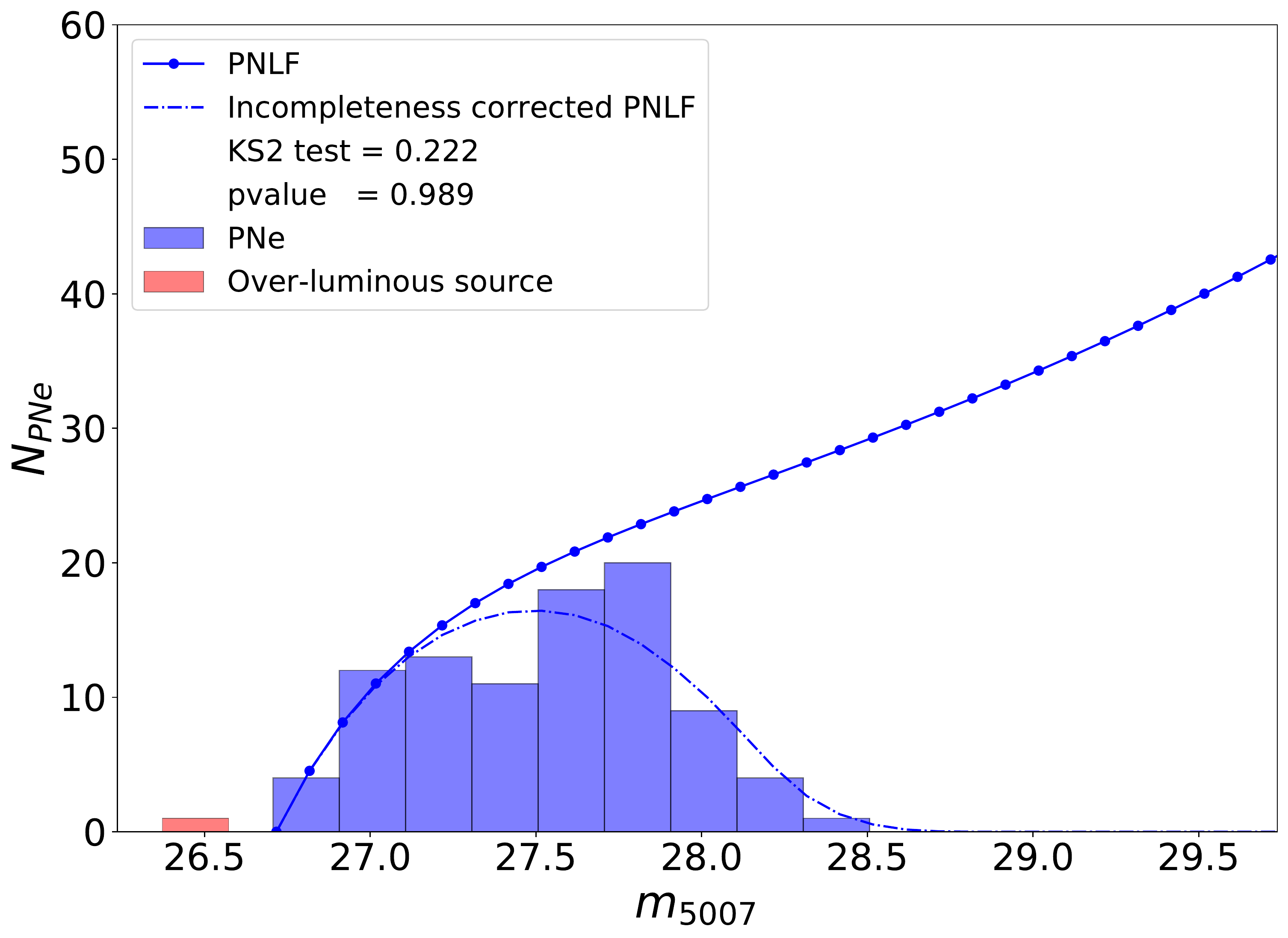}
    \caption{PNLF for FCC\,167 given by the binned values of $\rm m_{5007}$ from the PNe population. The blue solid line indicates the empirical form of the PNLF, given by Eq.~\ref{eq:PNLF}, and the completeness-corrected PNLF
    is depicted by the dashed line. The two curves are normalised such that the integral of the incompleteness-corrected PNLF matches the total number of observed PNe.}
    \label{fig:FCC167_PNLF}
    \end{figure}
}

\newcommand{\placefigPNLFb}{
    \begin{figure}
    \centering
    \includegraphics[width=\hsize]{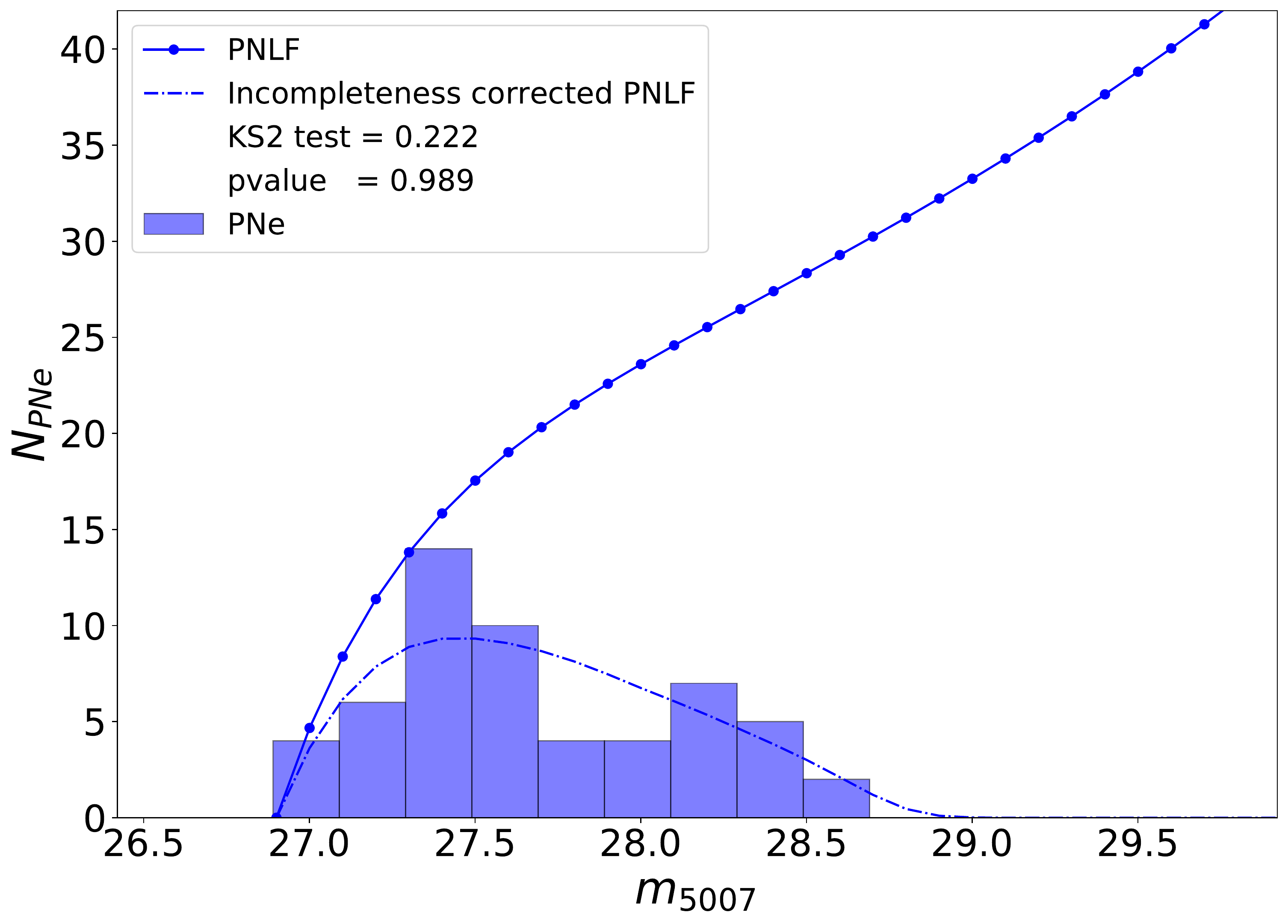}
    \caption{PNLF for FCC\,219, shown by the binned values of $\rm m_{5007}$ from the catalogued PNe. Similar to Fig.~\ref{fig:FCC219_A_rN}, we plot the empirical PNLF and the completeness corrected function.}
    \label{fig:FCC219_PNLF}
    \end{figure}
}

\begin{document}

   \title{Fornax 3D project: Automated detection of planetary nebulae in the centres of early-type galaxies and first results\thanks{Tables 4 and 5 are only available in electronic form
   at the CDS via anonymous ftp to cdsarc.u-strasbg.fr (130.79.128.5) or via \href{http://cdsarc.u-strasbg.fr/viz-bin/cat/J/A+A/637/A62}{http://cdsarc.u-strasbg.fr/viz-bin/cat/J/A+A/637/A62}}}

   \subtitle{}

   \author{T. W. Spriggs\inst{1}\thanks{TSpriggs@outlook.com}
   \and
   M. Sarzi \inst{1,2}
   \and
   R. Napiwotzki \inst{1}
   \and
   P. M. Gal\'an-de Anta \inst{2,3}
   \and
   S. Viaene \inst{1,5}
   \and
   B. Nedelchev \inst{1}
   \and
   L. Coccato \inst{8}
   \and
   E.\,M.\,Corsini \inst{11, 12}
   \and
   P. T. de Zeeuw \inst{6,7}
   \and
   J. Falcón-Barroso \inst{13, 14}
   \and
   D. A. Gadotti \inst{8}
   \and
   E. Iodice \inst{4}
   \and
   M. Lyubenova \inst{8}
   \and
   I.\,Martín-Navarro \inst{9,10}
   \and
   R. M. McDermid \inst{15}
   \and
   F. Pinna \inst{13}
   \and
   G. van de Ven \inst{16}
   \and
   L. Zhu \inst{10}
    }

   \institute{Centre for Astrophysics Research, School of Physics, Astronomy and Mathematics, University of Hertfordshire, College Lane, Hatfield AL10 9AB, UK 
    \and
    Armagh Observatory and Planetarium, College  Hill, Armagh BT61 9DG, Northern Ireland, UK 
    \and
    Astrophysics Research centre, School of Mathematics and Physics, Queen's University Belfast, Belfast BT7 INN, UK 
    \and
    INAF–Osservatorio Astronomico di Capodimonte, via Moiariello 16, I-80131 Napoli, Italy 
    \and
    Sterrenkundig Observatorium, Universiteit Gent, Krijgslaan 281, 9000 Gent, Belgium 
    \and
    Sterrewacht Leiden, Leiden University, Postbus 9513, 2300 RA Leiden, The Netherlands 
    \and
    Max-Planck-Institut fuer extraterrestrische Physik, Giessenbachstrasse, 85741 Garching bei Muenchen, Germany 
    \and
    European Southern Observatory, Karl-Schwarzschild-Strasse 2, D-85748 Garching bei Muenchen, Germany 
    \and
    University of California Observatories, 1156 High Street, Santa Cruz, CA 95064, USA 
    \and
    Max-Planck-Institut fuer Astronomie, Koenigstuhl 17, D-69117 Heidelberg, Germany 
    \and
    Dipartimento di Fisica e Astronomia ‘G. Galilei’, Università di Padova, vicolo dell’Osservatorio 3, I-35122 Padova, Italy 
    \and
    INAF–Osservatorio Astronomico di Padova, vicolo dell’Osservatorio 5, I-35122 Padova, Italy 
    \and
    Instituto de Astrof\'isica de Canarias, V\'ia L\'actea s/n, E-38205 La Laguna, Tenerife, Spain 
    \and
    Departamento de Astrof\'isica, Universidad de La Laguna, E-38205 La Laguna, Tenerife, Spain 
    \and
    Department of Physics and Astronomy, Macquarie University, Sydney, NSW 2109, Australia 
    \and
    Department of Astrophysics, University of Vienna, T\"urkenschanzstrasse 17, 1180 Vienna, Austria 
    }

   \date{Received 07/10/2019; accepted 20/03/2020}

    \titlerunning{Fornax 3D: Automated PNe detection survey in early-type galaxies}

\abstract
   {Extragalactic planetary nebulae (PNe) are detectable through relatively strong nebulous [\ion{O}{iii}] emission and act as direct probes into the local stellar population. Because they have an apparently universal invariant magnitude cut-off, PNe are also considered to be a remarkable standard candle for distance estimation.
   Through detecting PNe within the galaxies, we aim to connect the relative abundances of PNe to the properties of their host galaxy stellar population. By removing the stellar background components from FCC\,167 and FCC\,219, we aim to produce PN luminosity functions (PNLF) of these galaxies, and thereby also estimate the distance modulus to these two systems. Finally, we test the reliability and robustness of our novel detection and analysis method.
   It detects unresolved point sources by their [\ion{O}{iii}] 5007\AA{} emission within regions that have previously been unexplored. We model the [\ion{O}{iii}] emissions in the spatial and spectral dimensions together, as afforded to us by the Multi Unit Spectroscopic Explorer (MUSE), and we draw on data gathered as part of the Fornax3D survey. For each source, we inspect the properties of the nebular emission lines to remove other sources that might hinder the safe construction of the PNLF, such as supernova remnants and \ion{H}{ii} regions. As a further step, we characterise any potential limitations and draw conclusions about the reliability of our modelling approach through a set of simulations. By applying this novel detection and modelling approach to integral field unit observations, we report for the distance estimates and luminosity-specific PNe frequency values for the two galaxies. Furthermore, we include an overview of source contamination, galaxy differences, and possible effects on the PNe populations in the dense stellar environments.}

   \maketitle

%

\section{Introduction}

Planetary nebulae (PNe) originate in a spectacular event that occurs towards the end of the lifetime of most 2-8 $M_{\sun}$ stars, where copious amounts of oxygen-rich stellar material is expelled outwards. The ejected material is subsequently ionised by UV radiation from the central star, with the forbidden [\ion{O}{iii}] 5007 \AA{} line being prominent in many PNe, accompanied by the doublet line at 4959 \AA{}. Planetary nebulae thus act as isolated beacons within galaxies, allowing for their detection through spectroscopic observations \citep[e.g.][]{Paczynski1971EvolutionNebulae, Dopita1992AScale}.

The study of extragalactic PNe is centred around three major areas of research. The planetary nebular luminosity function (PNLF) is a viable distance indicator \citep{Ciardullo1989PlanetaryCompanions}. Furthermore, PNe can be used as direct probes of galaxy halo kinematics and dark matter content \citep{Romanowsky2003AGalaxies, Douglas2007The3379, Coccato2009KinematicNebulae, Kafle2018TheGalaxy, Longobardi2018KinematicsNebulae, Pulsoni2018TheMass, Bhattacharya2019TheNebulae}. Finally, PNe can be used to better understand the later stages of stellar evolution, and in particular, stellar environments different from those in our Galaxy \citep[e.g. stellar metallicity and kinematics;][]{Marigo2004EvolutionPNLF}.

One of the more traditional techniques of detecting extragalactic PNe, with their radial velocity, is so-called on-off band imaging, followed by spectroscopic measurements (either multi-slit or slitless spectroscopy). Counter-dispersed slitless spectroscopy, used for example in the Planetary Nebulae Spectrograph \citep[][an instrument that is entirely dedicated to the study of extragalactic PNe]{Douglas2002TheKinematics}, offer a better solution because it is capable of identifying and measuring position and velocity with a single observation, without follow-ups. In all these techniques, the identification of extragalactic PNe is limited to the halo and outer regions of the host galaxy (typically > 0.5 effective radii, $R_e$), where the stellar continuum background does not dominate. In this way, PNe have been detected within the intra-cluster mediums of the Coma and Hydra clusters \citep{Gerhard2005DetectionCluster, Ventimiglia2011The1060}, which reside at 100 Mpc and 50 Mpc, respectively.

Previously, extragalactic surveys such as the Spectrographic Areal Unit for Research on Optical Nebulae (SAURON) survey \citep{Bacon2001TheSpectrograph} on the William Herschel Telescope (WHT), the Calar Alto Legacy Integral Field Area (CALIFA) survey \citep{Sanchez2011CALIFAPresentation} on the CAHA telescope \citep{Roth2005PMASPerformance}, and the Mapping Nearby Galaxies at Apache Point Observatory (MANGA) survey \citep{Bundy2014OVERVIEWOBSERVATORY} on the Sloan Digital Sky Survey (SDSS) \citep{York2000TheSummary} have shown that modelling the background stellar continuum within galaxies is feasible and can be applied to a variety of data sets. This allows us to cleanly isolate the ionised-gas emission in the galaxy spectra and to map the nebular activity across the entire field of view, including that originating from unresolved PNe sources. In this respect, the studies of \citet{Sarzi2011TheView} and \citet{Pastorello2013TheView} based on SAURON data for M32 and the central regions of Andromeda clearly illustrate the ability of integral field spectroscopy of detecting PNe down to the very central regions of external galaxies. With the Multi Unit Spectroscopic Explorer (MUSE) \citep{Bacon2010TheInstrument}, we can detect PNe at even farther distances, in particular because its collecting power and spatial resolution are superios. This was illustrated by \citet{Kreckel2017AMUSE} in the case of the spiral galaxy NGC~628, which lies at a distance of 9.6 Mpc. At twice the distance, \citet{Sarzi2018AstronomyResults} later presented preliminary results for one of the two Fornax cluster galaxies that are covered by the present study (FCC\,167). Adaptive optics MUSE observations (e.g. \citet{Fahrion2019ConstrainingFCC47}) will certainly push the detections of extragalactic PNe even further.

With the detection of PNe by their prominent [\ion{O}{iii}] 5007\,\AA{} emissions, we can then start to investigate and catalogue their characteristics, which are total [\ion{O}{iii}] flux, apparent magnitudes, emission line ratios, and line-of-sight velocities. Starting with the relative [\ion{O}{iii}] 5007\AA{} apparent magnitude, the flux of a given PNe can be converted into a V-band corrected magnitude \citep[][Eq.~\ref{eq:m5007}]{Ciardullo1989PlanetaryCompanions},
\begin{equation}
    m_{5007} = -2.5 \log_{10}(F_{[\ion{O}{iii}} \ (\mathrm{erg \ cm^{-2} \ s}^{-1})) - 13.74.
    \label{eq:m5007}
\end{equation}
When sources of [\ion{O}{iii}] emissions have been identified and confirmed as PNe, we can then produce a PNLF for our detected sample and compare its shape to an empirically derived functional form of the PNLF reported by \citet{Ciardullo1989PlanetaryCompanions}, which we discuss in Sect.~5. Estimating the PNLF is intrinsic to the process of using PNe as standard candle estimators. In the extragalactic context, the PNLF has been shown to exhibit a cut-off value towards the bright end. Under the assumption that a universal PNLF holds true for all galaxies and that its brightest PNe are indeed detected, we can use the conversion of apparent into absolute magnitudes to obtain an estimate for the distance to the host system. The steps of the analysis with which we derive this estimate are presented in Sect.~5. The luminosity-specific PN frequency, referred to as the $\alpha$ value, is another measurement that is based on the PNLF. It is a proxy for the number of PNe that are expected to be produced per unit stellar luminosity of a particular galaxy.

Previous works have reported several interesting correlations of the halo $\alpha$ values and intrinsic host galaxy properties \citep{Buzzoni2006PlanetaryPopulations, Coccato2009KinematicNebulae, Cortesi2013TheKinematics}. \citet{Buzzoni2006PlanetaryPopulations} showed that $\alpha$ appears to be connected to the host galaxy metallicity and UV excess. They found that as the core UV excess of the host galaxy increased, the $\alpha$ value of the halo decreased. A similar correlation was also found when the core metallicity was compared to the $\alpha$ value of the halo. This was interpreted to stem from the effects metallicity has on stellar evolution and subsequent PNe formation. Further examples of reported correlations include the kinematics of the PNe system (parametrised either with the root mean square (rms) velocity ($V_{rms}$) or the shape of the velocity dispersion radial profile) with galaxy luminosity (optical and X-ray), angular momentum, isophotal shape parameter, total stellar mass, and the $\alpha$ parameter \citep{Coccato2009KinematicNebulae}.

Within this context, it is important to note that whereas our knowledge of the shape and normalisation of the PNLF comes chiefly from peripheral PNe populations of galaxies, measurements for the stellar metallicity and the UV spectral shape of the galaxies typically pertain to their central regions (well within one Re). For instance, \citet{Buzzoni2006PlanetaryPopulations} compared the properties of halo PNe populations with central measurements for the stellar kinematics, metallicity, and UV excess.
Integral-field spectroscopy (IFS) can overcome this spatial inconsistency because IFS not only enables revealing PNe deeply in the central regions of galaxies, but measuring the stellar age and metallicity in the same regions in which the PNe are detected, including regions such as the stellar halos  (\citet[first illustrated by][]{Weijmans2009Stellar821}).

To progress along these lines, in this paper we illustrate that with the aid of IFS, we can characterise the PNe populations of external galaxies on the basis of the MUSE observations for two early-type galaxies in the Fornax cluster, FCC\,167 (NGC 1380) and FCC\,219 (NGC 1404). When the central cluster galaxy NGC~1399 is excluded, these are the two brightest objects inside the virial radius of the Fornax cluster, with a total r-band magnitude of $m_r=9.3$ and $8.6$, respectively \citep{Iodice2019TheVST}. These ETGs are morphologically and dynamically different, however. FCC\,167 is a fast-rotating S0a galaxy, whereas FCC\,219 is a slowly rotating E2 galaxy with a kinematically decoupled core (\citealp{Ricci2016IFUKinematics}, but see also \citealp{Iodice2019TheMaps}). Furthermore, FCC\,219 is known to host a substantial hot-gas halo (e.g. \citealp{Machacek2005InfallCluster}), whereas FCC\,167 shows a much weaker X-ray halo in the deep XMM-Newton images from \citet{Murakami2011SuzakuDistribution} and \citet{Su2017GasCluster}. This difference is relevant to the discussion of our PNe results for these two objects. For this paper, we initially assume a distance of 21.2 Mpc and 20.2 Mpc for FCC\,167 and FCC\,219, respectively, according to the surface-brightness fluctuation measurements of \citet{Blakeslee2009THEDISTANCE}.

The paper is structured as follows: Section 2 details the data and targets from the Fornax 3D survey, along with a brief overview of the data analysis steps that aided our detection of PNe. Section 3 introduces our method regarding any of the PNe [\ion{O}{iii}] measurement and fitting processes, as well as the steps we used in object identification. It also describes our PNe modelling simulations and our procedure for obtaining a sensible estimate of the instrumental point spread function (PSF). In Section 4 we describe the results we obtained by running our novel 1D+2D modelling approach, which is capable of successfully identifying and extracting the PNe populations of FCC\,167 and FCC\,219. Section 6 gives a brief overview of the PNLF, and we describe how we accounted for incompleteness. Finally, we compare our results with similar previous studies in Sect. 3 and discuss the reasons for the potential difference between some of the available distance estimators and our estimates.

\section{Observations and data reduction}

FCC\,167 and FCC\,219 were observed with the MUSE IFS unit \citep{Bacon2010TheInstrument} as part of the magnitude-limited survey of galaxies within the virial radius of the Fornax galaxy cluster \citep[hereafter Fornax3D]{Sarzi2018AstronomyResults}.
To cover their central and outer regions (down to a surface-brightness limit of $\mu_B=25\ \textrm{mag} \ \textrm{arcsec}^{-2}$), wide-field mode seeing-limited MUSE data were acquired over three and two separate pointings for FCC\,167 and FCC\,219, respectively, with total exposure times of 1h for central pointings and 1h30m for the intermediate or outer pointings. This provides high-quality spectroscopic measurements in 0.2\arcsec $\times$ 0.2\arcsec\ spatial elements over a 4650--9300\AA{} wavelength range, with a spectral sampling of 1.25\AA{} $\textrm{pixel}^{-1}$.
As detailed in \citeauthor{Sarzi2018AstronomyResults}, our MUSE data were reduced using the MUSE pipeline \citep{Weilbacher2012DesignPipeline, Weilbacher2016MUSE-DRP:Pipeline} within the ESOREFLEX \citep{Freudling2013AutomatedAstronomy} environment, where special care was taken to remove the sky background through the use of dedicated sky field exposures and of the Z\"urich atmospheric purge algorithm \citep{Soto2016ZAPSpectroscopy}.

For the purpose of this paper, we obtained final datacubes for each pointing without further combining these into a single mosaic (as shown e.g. in the case of FCC\,167 in \citealp{Sarzi2018AstronomyResults}). To enable the study of galactic nuclei and unresolved sources such as PNe and globular clusters, the central pointings had a stricter imaging requirement (FWHM < 1.0\arcsec) than the intermediate or outer pointings (FWHM < 1.5\arcsec) during the Fornax3D observations. In this way, we used the data with the highest MUSE quality in the regions in which our pointings overlap spatially.

Finally, to ensure that the absolute flux calibration of our datacubes is correct, we applied the same procedure to FCC219 as was used in the case of FCC\,167 in \citet{Sarzi2018AstronomyResults} to compare this with images obtained with the Hubble Space Telescope (HST). Similarly, the MUSE flux densities for FCC\,219 closely match those of the HST.

\section{PNe sources identification and flux measurements}

To compile a catalogue of PNe in our two target galaxies and measure their [\ion{O}{iii}] flux values, we first proceed with a dedicated reanalysis of the MUSE datacubes (Sect.~3.1), then draw a conservative list of PNe candidate sources (Sect.~3.2), and finally fit (Sect.~3.3) and validate (Sect.~3.4) each PNe candidate with a 1D+2D model\footnote{\href{https://github.com/tspriggs/MUSE_PNe_fitting}{GitHub/MUSE\_PNe\_fitting} \citep{Spriggs2020Tspriggs/MUSE_PNe_fitting:Release}} that accounts for the expected unresolved spatial distribution of the [\ion{O}{iii}] flux while optimising for the spectral position of the [\ion{O}{iii}] lines. A prior evaluation of the spatial PSF is needed to inform this final fit (Sect.~3.5), which is done either using foreground stars or by simultaneously applying our 1D+2D-model fitting approach to several bright PNe sources.

\subsection{Isolating the nebular emission component}

To identify and fit PNe sources, we used pure emission-line datacubes that we obtained after evaluating and subtracting the stellar continuum from each individual MUSE spectrum in our pointing datacubes. As detailed in \citet{Sarzi2018AstronomyResults} and also shown in \citet{Viaene2019The167}, this was done through a spaxel-by-spaxel simultaneous fit for the stellar and ionised-gas spectrum using the GandALF \citep{Sarzi2006TheGalaxies} fitting tool, which in turn is informed by previous fits with pPXF \citep{Cappellari2004ParametricLikelihood, Cappellari2017ImprovingFunctions} and GandALF on Voronoi-binned spectra \citep{Cappellari2003AdaptiveTessellations}, drawing on the IFU data-processing pipeline of \citet[GIST\footnote{\href{https://abittner.gitlab.io/thegistpipeline}{https://abittner.gitlab.io/thegistpipeline}}][]{Bittner2019TheData}.

While emission-line cubes and other pipeline (stellar fitting) results are in principle available from this analysis \citep[see also][]{Iodice2019TheMaps}, in the case of extended targets such as FCC\,167 and FCC\,219, we repeated our fitting procedure for individual MUSE pointing datacubes. Furthermore, to achieve the best-fit quality, and as described in \citet{Sarzi2018AstronomyResults}, the entire MILES \citep{Sanchez-Blazquez2006Medium-resolutionSpectra} stellar library was used to match the stellar continuum rather than resorting to stellar population models (as shown in Fig. 5 of \citealp{Sarzi2018AstronomyResults}). This is necessary to minimise stellar contamination, which might affect our scientific goals, and it improved the reliability of our nebular emission extraction.

\placefigone
\placefigtwo

\subsection{Identification of PNe candidates}
To obtain an initial list of PNe source candidates, we drew from our spaxel-by-spaxel [\ion{O}{iii}]\,5007\,4959 \AA\AA{} line-fit results. Although our GandALF fits also provide this information, these fits are not properly optimised for the detection of PNe. We used GandALF to capture the general behaviour of any present ionised gas, including some regions of diffuse ionised-gas emission or AGN activity. These were safely identified and isolated from any potential unresolved PNe emission. When PNe are to be located, it is better to explicitly account for the fact that PNe have only modest expansion velocities \citep[between 10 and 40 $ \rm km\,s^{-1}$;][]{Weinberger1983AAge, Hajian2007AnNebulae, Schonberner2014AObservation}, which leads to [\ion{O}{iii}] line profiles that are probably near the instrumental resolution limit. For this reason, we refit for the [\ion{O}{iii}]\,4959 and 5007 \AA\AA\  doublet in the 4900 -- 5100\,\AA\ spectral region of our pure emission-line cubes. Here, we assumed a constant intrinsic stellar velocity dispersion for the PN (see Sect. 3.4 below for values) and an instrumental spectral resolution ($\sigma_{\mathrm{MUSE,LSF}}$) at 5007\AA\ of 75 $\rm km\,s^{-1}$ , according to the MUSE line-spread function (LSF) behaviour as measured by \citet{Guerou2017TheSurvey}.

The results of this dedicated [\ion{O}{iii}] doublet fit and its ability to reveal unresolved [\ion{O}{iii}] sources is shown in Figs. \ref{fig:FCC167_A_rN} and \ref{fig:FCC219_A_rN} for FCC\,167 and FCC\,219, respectively. In particular, these maps show the value for the ratio of the fitted peak amplitude of the [\ion{O}{iii}]\,5007 line ($A$) and the residual-noise level ($rN$) from our fits around the [\ion{O}{iii}] doublet. The residual noise level was evaluated as the standard deviation of the residuals after subtracting our [\ion{O}{iii}] model from the data. As discussed in \citet{Sarzi2006TheGalaxies}, this $A/rN$  is a good measure for the threshold beyond which emission lines can be detected and for how well they can be measured. These $A/rN$ maps therefore provide a better contrast between [\ion{O}{iii}] sources and regions dominated by false-positive [\ion{O}{iii}] detection than maps of either line amplitudes or fluxes.

With these signal-to-noise ratio ($A/rN$) maps at hand, we compiled an initial list of PNe candidates using the Python package SEP. This is a python script-able version of the popular Sextractor source-finding routine of \cite{Bertin1996SExtractor:Extraction}. First, background noise is evaluated, and the subsequent noise map is subtracted from the $A/rN$ data. Having first tested this approach against a no-background-subtraction attempt that also used a larger central exclusion region, we concluded that subtracting the SEP-derived background aided avoiding spurious sources, and meant that the masked region could be much smaller than before.
We adopted a rather conservative source-detection threshold corresponding to two standard deviations above the SEP derived, background noise. This decision proved to strike the balance between detecting an excessive number of sources, which would result in a larger subsequent fraction of source exclusions, versus detecting the more prominent sources, which would make up the majority of the validated PNe. Figures \ref{fig:FCC167_A_rN} and \ref{fig:FCC219_A_rN} highlight the sources that are present within the field of view (FOV). Excluded sources are circled in red. The dashed line regions are excluded because they include known diffuse ionised-gas emission (e.g. for FCC\,167 in Fig.~\ref{fig:FCC167_A_rN}, see also \citealp{Viaene2019The167}), or contain regions where a template-mismatch was found to bias our flux measurements (e.g. seen in FCC\,219, in Fig.~\ref{fig:FCC219_A_rN}).

\placefigthree

\subsection{Candidate PNe fitting}
To validate the unresolved nature of our PNe candidates and measure their kinematics and total [\ion{O}{iii}] 5007\AA{} fluxes, we started by fitting each of them with a procedure that used all the information contained in the emission-line cube, near the spatial location of the source (in a $9\times9$ spaxel region that is 1.8\arcsec across) and in the wavelength region around the [\ion{O}{iii}] doublet (between 4900\,\AA{} and 5100 \AA). Specifically, we simultaneously matched all the MUSE spectra in this portion of our emission-line cube with a 1D+2D emission-line model where the total [\ion{O}{iii}] 5007\AA{} flux ($F_{\rm [\ion{O}{iii}]}$) of the model, the shape of the PSF, and the exact spatial positioning of the model determines the [\ion{O}{iii}] model flux at each spaxel ($F_{\rm [\ion{O}{iii}]} (x, y)$). This in turn can be translated into a model [\ion{O}{iii}] 5007 4959 \AA\AA{} line profile through our knowledge of instrumental LSF and the optimisation of the PN emission intrinsic width $\mathrm{\sigma_{PNe}}$ and velocity $v$ (although in practice we solved for the total profile width $\mathrm{\sigma_{tot}}$, which incorporates the convolution of the LSF and $\mathrm{\sigma_{PNe}}$).

When we assume a \citet{Moffat1969APhotometry} profile for the PSF, the [\ion{O}{iii}] 5007\AA{} flux distribution around a PNe source can be written as
\begin{equation}
    F_{[\rm \ion{O}{iii}]}(x,y) = F_{\rm [\ion{O}{iii}]}(x_0,y_0)\ \bigg(1 + \frac{( x - x_0 )^2 + ( y - y_0)^2 }{\alpha^2} \bigg)^{-\beta},
    \label{eq:Moff}
\end{equation}
where $\alpha$ and $\beta$ determine the radial extent and kurtosis of the Moffat distribution, $x_0$ and $y_0$ locate the source centre, and $F_{\rm [\ion{O}{iii}]} (x_0,y_0)$ is the peak [\ion{O}{iii}] flux at this position. The latter is related to the total [\ion{O}{iii}] flux through the Moffat profile normalisation
\begin{equation}
    F_{\rm [\ion{O}{iii}]}(x_0,y_0) = F_{\rm [\ion{O}{iii}]}\;\frac{\beta-1}{\pi\alpha^2}.
\end{equation}
The spatial extent of our sources can also be quantified using the FWHM of the Moffat profile, which is given by
\begin{equation}
    {\rm FWHM} = 2 \alpha \sqrt{2^{1 / \beta}-1}.
    \label{eq:FWHM}
\end{equation}
In its more general form, this model therefore includes seven free parameters ($F_{\rm [\ion{O}{iii}]}, v, \mathrm{\sigma_{total}}, x_0, and y_0$ for the PN source, and $\alpha$ and $\beta$ for the PSF) plus two additional parameters to account for background that remains after the continuum subtraction (spectrum background level and gradient). These parameters are all optimised through a standard non-linear $\chi^2$ minimisation \citep{Newville2014LMFIT:Python, Newville2019Lmfit/lmfit-py1.0.0}.
In practice, however, the full set of parameters is only varied initially, when the PSF of our observations is constrained or when the typical value of $\mathrm{\sigma_{tot}}$ (Sect.~3.5) is estimated. After the values of $\mathrm{\sigma_{tot}}$ and the PSF were determined, we held $\mathrm{\sigma_{tot}}$, $\alpha$, and $\beta$ fixed and allowed the remaining parameters to vary.

Figure~\ref{fig:FCC219_spaxel_by_spaxel} illustrates the working of our 1D+2D-fitting in the case of a PN source (61) in the central region of FCC\,219 (Fig.~\ref{fig:FCC219_A_rN}). In particular, by imposing a PSF behaviour on the intensity of the model profile for the [\ion{O}{iii}] doublet at each spaxel position around the candidate PN, our approach automatically checks the unresolved nature of the emission-line source while minimising any bias on the recovered parameters that might be introduced by regions with little or no [\ion{O}{iii}] flux. Furthermore, by also considering the spectral component of the data, this technique allows us to isolate PNe that are embedded in diffuse ionised-gas components, as well as to distinguish two blended PNe sources with different kinematics \citep[see e.g.][]{Pastorello2013TheView}. Because PNe are ubiquitous, this method finally also offers a way to measure the PSF when galaxies are targeted with IFS observations.

\placefigfour
\placefigfive

\subsection{PNe candidate validation}

When we had our initial 1D+2D fits, we were able to further filter objects that did not show typical PNe characteristics. They include detections that are inconsistent with unresolved sources, in which our fit results might either be biased by broader ionised-gas emission or have a fair chance of being the consequence of a false-positive detection.

To verify that the [\ion{O}{iii}] flux distribution of our candidate source was consistent with a given PSF, we relied on the quality of our fits. We therefore excluded objects for which the $\chi^2$ value returned by our fit was outside the 95\%\  confidence limit for a $\chi^2$ distribution with $\nu$ degrees of freedom (corresponding to $9 \times 9 \times N_{\lambda} $ data points, minus the six free parameters).

To understand when our fit results can be deemed reliable, we ran a set of simulations in particular to determine when the recovery of key parameters becomes biased in the regime of a low signal-to-noise ratio.
For this, we created a number of mock PNe data in the same type of emission-line $9\times9$ minicubes that were passed to our 1D+2D-fitter. The total $F_{\rm [\ion{O}{iii}]}$ values corresponded to PNe over a range of absolute $\rm M_{5007}$ magnitudes (between -4.5 and -1.0 in steps of 0.05), observed at the estimated $\mathrm{D_{PNLF}}$ in FCC\,167 (see Sect.~5). These emission lines were spatially distributed according to the measured PSF of our central observations, where the peak of the emissions was located at towards the centre of the minicube. The local [\ion{O}{iii}] flux at each spaxel was then converted into an [\ion{O}{iii}] doublet profile for a PNe moving at the systemic velocity of FCC\,167 ($1878 \ \rm km s^{-1}$ \citet{Iodice2019TheMaps}), the [\ion{O}{iii}] profile width was set to the value of $\mathrm{\sigma_{tot} \sim 100 \ \rm km\,s^{-1}}$, as found from the PSF fitting (Sect.~3.5), which corresponds to an intrinsic line broadening ($\mathrm{\sigma_{PNe}\ of \ 40 km\,s^{-1}}$), considering the value of the MUSE LSF ($\mathrm{\sigma_{MUSE,LSF}}$). Finally, random Gaussian noise was added to a level typical of the fit-residual noise (rN) found in the central pointing of FCC\,167 (located outside the masked region of Fig.\ref{fig:FCC167_A_rN}, predominantly containing diffuse ionised-gas emission). This simulation set-up allowed us to explore how well our 1D+2D-fitting approach recovered the total flux of the simulated PNe, together with the flux distribution and kinematics, at different levels of background-noise contamination. To quantify the latter, we considered the ratio between the maximum [\ion{O}{iii}] 5007 line amplitude at the centre of our model and the residual noise level ($rN$).
\begin{equation}
    \sigma_{tot} = \sqrt{\sigma_{MUSE, LSF}^{2} + \sigma_{PNe}^{2}},\end{equation}
\begin{equation}
    A_{\rm [\ion{O}{iii}]}(x_0,y_0) =\frac{F_{\rm [\ion{O}{iii}]}(x_0,y_0)}{\sqrt{2 \pi \,} \sigma_{tot}}.
\end{equation}
Starting with the most general case (i.e. trying to recover all parameters), Fig.~\ref{fig:free_PSF_sims} shows not only that the accuracy of the recovered parameters decreases at lower central $A/rN$ values, but it also highlights parameters that are biased towards the regime of a lower signal-to-noise ratio. These biases appear most pronounced in the PSF parameters, notably below five A/rN, along with the estimation of the [\ion{o}{iii}] emission velocity dispersion $\sigma_{tot}$ of the PN. Although these biases have only a limited knock-on effect on the measured total flux, at low $A/rN$ values, the uncertainties in the absolute $\rm M_{5007}$ magnitudes quickly reach values that can affect the distance estimates (e.g. a 0.2 magnitude error implies a 10\%\ error on any attempted distance estimates based on the PNLF).

As mentioned above, whereas Fig.~\ref{fig:free_PSF_sims} has some relevance for our PSF determination (see Sect.~3.5), we typically measured our PNe candidate sources while holding to the best-fit PSF parameters and value of $\sigma$ derived from bright PNe or stars in the field. Under these conditions, the behaviour of our 1D+2D-fitting approach is shown by Fig.~\ref{fig:fixed_PSF_sims}, by which we can conclude that above a central $A/rN=3,$ the recovered absolute magnitude values are essentially unbiased and accurate to 0.1 magnitude or below.

\placefigsix

To complement these idealised simulations, we also assessed the level of false-positive contamination by running our 1D+2D-fitting code at randomly selected locations. In this procedure, we excluded regions with known diffuse ionised-gas emission, as well as the locations of our candidate PNe sources. Figure~\ref{fig:FCC167_falsepositive} shows the distribution for the central $A/rN$ values obtained from fits to noise for FCC\,167, indicating that 99\% of the amplitude of these false-positive ($\mathrm{A_{false\ pos}}$) results lie below three times the residual noise ($A_{\mathrm{false\ pos}}/rN <3$). The grey lines in Fig.~\ref{fig:FCC167_falsepositive} illustrate the results of fitting sources that lie closer to the central masked regions of the galaxy because of the higher complexity and density of stellar light. Here, template mismatch produces an erroneous [\ion{O}{iii}] signal, that is, increased background levels, which causes the fitter to mistake spectral noise for [\ion{O}{iii}] emission lines. This then produces higher $A_{\mathrm{false\ pos}}/rN$ values, with a greater fraction of $A_{\mathrm{false\ pos}}/rN$ above our imposed cut-off of three times the residual noise.

In summary, we validated our PNe candidate sources using standard $\chi^2$ statistics to verify the quality of our 1D+2D fits and considered only objects where the central $A/rN > 3$. We excluded regions with diffuse ionised-gas emission or in which template mismatch can lead to a false-positive detection with the central $A/rN$ above this threshold.

\subsection{PSF determination}

An accurate knowledge of the PSF is key to our PNe flux measurements. A Moffat (1969) profile (Eq.~\ref{eq:Moff}) generally describes the PSF of astronomical observations well, including those obtained with MUSE \citep{Bacon2010TheInstrument}.

To measure the PSF from our MUSE data, we relied either on foreground stars in the FOV of our MUSE pointing or on the PNe sources themselves when no star was available.
For these two situations, we modified on the one hand our 1D+2D-fitting code, which allowed us to ignore the spectral direction and thus fit the flux distribution of a star, and on the other hand, we implemented the option to fit several PNe sources at the same time.
In the latter case we used the same PSF parameters $\alpha$ and $\beta$ and a common intrinsic $\mathrm{\sigma_{tot}}$ for the different sources, while we individually optimised only for $F_{\rm [\ion{O}{iii}]}, v, x_0, y_0$ and for the two continuum-shape nuisance parameters: gradient and background level. Typically, we found that it was enough to consider up to ten of the better detected PNe during this process These PNe had central $A/rN$ values of at least eight, as estimated from an initial 1D+2D-fit with all parameters free to achieve a satisfactory estimate for the PSF. When the PSF from foreground stars was constrained, we also allowed for a constant flux background from the host galaxy.

\placefigseven

To illustrate the accuracy with which the PSF is measured using PNe sources, we show in Fig.~\ref{fig:PSF_star_vs_PNe} the surface-brightness profile for the foreground southwestern star in the FOV of the central pointing of FCC\,219 and the associated star that fits a Moffat model best. We note that it compares rather well to the best-fitting Moffat profile as extracted from PNe. This is further quantified in Table~\ref{tab:PSF}, where we compare our PSF estimate with the one provided by the MUSE cube, as determined on the basis of a fit to the galaxy itself. Typical errors in the PSF parameters translate into total PSF flux uncertainties of less than 9\%. In particular, with this accuracy for the PSF, we are able to set a limit of the potential systematic error on our PNe magnitude estimates smaller than 0.1 mag.
To conclude, we note that even when a star was present, we ran our simultaneous PNe procedure in order to constrain the typical $\sigma_{tot}$ of the PNe in the target galaxy.

\begin{table}
    \renewcommand{\arraystretch}{1.2}
    \caption{Best estimates for the Moffat PSF parameters in our target galaxies, as derived using either foreground stars or PNe. The corresponding FWHM of our Moffat models are also compared to the FWHM measurements from the MUSE data header, as obtained by the MUSE slow-guiding system.} \label{tab:PSF}
    \begin{tabular}{l l l l l l l}
    \hline
    Galaxy & Method & $\alpha$ & $\beta$ & FWHM & $\rm FWHM_{HDR}$\\
    & & & & (pixels) & (pixels) \\
    \hline
    FCC 167 & PNe  & 2.99 & 2.15 & 3.69 & 3.56 \\
    FCC 219 & PNe  & 4.26 & 3.37 & 4.07 & 3.50 \\
    FCC 219 & Star & 4.29 & 3.42 & 4.07 & 3.50 \\
    \hline
    \end{tabular}

\end{table}

\placefigeightA
\placefigeightB

\subsection{Literature comparison}
To date, the most comprehensive PNe cataloguing for FCC\,167 and FCC\,219 was compiled by \citet{Feldmeier2007CalibratingResults} and \cite{McMillan1993PlanetaryCluster}, respectively. The PNe detection method used in these surveys is the "on-off" photometry because these studies find PNe mostly in the galaxy outskirts. We are however, able to match a select few sources located towards the edges of our central observations as well as in the outer pointings of each galaxy.

Within the central and disk pointings of FCC167, we matched 21 PNe with the records of \citet{Feldmeier2007CalibratingResults}, the majority of which are located outside of our central pointing. From comparing their catalogue, we conclude that we did not miss any PNe within our FOV. After comparing the measured magnitudes, we find a linear agreement that we show in Fig.\ref{fig:FCC167lit}, see Sect.~5.3, but with a systematic offset of 0.45 mag fainter than their recorded values. The origin of this offset is unclear. We are confident in our own flux calibration, which is based on HST images, and further note that Feldmeier's brighter $m_{5007}$ values lead to a rather small distance modulus for FCC167: $31.04^{+0.11}_{-0.15}$ (16.1 Mpc).

\placefiglita

For FCC219, \citet{McMillan1993PlanetaryCluster} reported nine PNe sources within the regions we mapped. One of their sources is excluded from our catalogue because it was filtered out. Fig.~\ref{fig:FCC219lit} shows the scatter of McMillan's $m_{5007}$ values versus those presented here. We also note that \citet{McMillan1993PlanetaryCluster} reported a distance modulus of $31.15^{+0.07}_{-0.1}$ (17.0 Mpc), which is $\sim$ 2 Mpc closer than our distance estimation.

Table~\ref{table:lit_comp} contains the object IDs of the matched PNe for FCC167 and FCC219 with their respective $m_{5007}$ from our measurements and those catalogued in \citet{Feldmeier2007CalibratingResults} and \citet{McMillan1993PlanetaryCluster}, respectively. We applied a separation limit of 3.6 arcseconds for matching sources. However, within FCC\,167, we had to account for a -0.4 to -0.8 arcsecond shift in declination coordinates, which may arise from a minor inaccuracy in the header information.

\placefiglitb

\begin{table}
    \renewcommand{\arraystretch}{1.2}
    \centering
    \caption{List of matched source IDs from the central pointings, accompanied by our measured $m_{5007}$ and those reported in the literature; FCC\,167: \citet{Feldmeier2007CalibratingResults}, and FCC\,219: \citet{McMillan1993PlanetaryCluster}.}
    \label{table:lit_comp}
    \begin{tabular}{llll}
    \hline
    Galaxy & ID & $m_{5007 \, F3D}$ & $m_{5007 \, lit}$\\
    \hline
    FCC167 & F3D J033627.54-345759.28 & 27.18 & 26.59 \\
           & F3D J033628.01-345814.80 & 26.92 & 26.73 \\
           & F3D J033626.37-345829.46 & 27.23 & 26.81 \\
           & F3D J033625.64-345818.91 & 27.61 & 27.20 \\
    \hline
    FCC219 & F3D J033849.09-353523.23 & 27.67 & 26.79 \\
           & F3D J033848.97-353520.76 & 27.08 & 26.83 \\
           & F3D J033850.08-353515.62 & 27.39 & 26.98 \\
           & F3D J033853.81-353502.60 & 28.23 & 27.54 \\
           & F3D J033849.53-353502.86 & 27.39 & 27.68 \\
    \hline
    \end{tabular}

\end{table}

\section{Spectral catalogue}

After selecting a robust set of PNe candidate sources, we proceeded to further characterise their spectral properties. Some of these unresolved [\ion{O}{iii}] sources may still originate from objects other than PNe. Typical PNe spectra are dominated by strong [\ion{O}{iii}] lines and little emission from other atomic species. Strong $\rm H\alpha$ emission, on the other hand, might signal unresolved H{\sc II} regions, whereas the additional presence of significant [\ion{N}{ii}] or [\ion{S}{ii}] emission might be indicative of a supernova remnant (SNR). A more comprehensive emission line fit might also contain information on the amount of extinction (through the Balmer decrement) and therefore lead to de-reddened absolute $M_{5007}$ magnitude values for our target PNe, which can fall below the PNLF cut-off value \citep[$M^{*}_{5007}=-4.53$][]{Ciardullo2012TheGaia}.

To make full use of the MUSE spectral range, we extracted a PSF-weighted MUSE aperture spectrum from the original MUSE cube at the location of our sources for each of our confirmed candidate
PNe sources. We then fit each of these aperture spectra using GandALF, keeping to the local stellar kinematics as derived in Sect.~3.1 and imposing the same profile on all emission lines. In particular, we fixed the width of all lines according to the value of the intrinsic $\sigma$ derived in Sect.~3.5.

Figures~\ref{fig:gandalf_PNe} and \ref{fig:gandalf_SNR} show two examples of these GandALF fits, one for the typical spectrum of a PNe source, and the other showing an example of a potential SNR impostor (Sect.~4.2). From these fits we obtained the fitted flux of H$\beta$, [\ion{O}{iii}]\,5007, [\ion{N}{ii}]\,6583, $\rm H\alpha,$ and of the [\ion{S}{ii}]\,6716,6731 doublet, together with their corresponding $A/rN$ values. We report a good agreement between the [\ion{O}{iii}] flux values with our 1D+2D fitting method from fits to these apertures. The values differed by less than 10\%.

\placefigvela
\placefigvelb
\placefigcontaminationa
\placefigcontaminationb

\subsection{PNe candidate interlopers}

We attempted to identify interlopers by determining whether the velocity of each PN candidate velocity, measured by the [\ion{O}{iii}] lines, was consistent with having been drawn from the local stellar line-of-sight velocity distribution (LOSVD). In Fig.~\ref{fig:FCC167_vel_hist} we plot the distribution for the ratio between the difference in the PNe candidate and local stellar velocity ($\mathrm{\Delta V = V_{PNe}-V_{stars}}$) and the local stellar velocity dispersion ($\mathrm{\sigma_{stars}}$). To first approximation, without accounting for higher-order moments of the LOSVD, we indeed expect this $\mathrm{\Delta V/\sigma_{stars}}$ ratio to follow a Gaussian distribution for PNe candidates that belong to the galaxy. Any object with $\mathrm{|\Delta V/\sigma_{stars}|>3}$ is very likely an interloper. One PNe is identified as interloping within FCC\,167, and is labelled as such in Table 4. We find no interloping PNe for FCC219. For FCC\,219, we plot the distribution of $\mathrm{|\Delta V/\sigma_{stars}|}$(see Fig.~\ref{fig:FCC219_vel_hist}). The distribution of the PNe and their respective velocities lies within the measured velocity distribution range reported in \citet{Iodice2019TheMaps} for both galaxies.

\subsection{PNe impostors}

It would be reasonable to assume that most of the unresolved point sources in an early-type galaxy that are detected through their [\ion{O}{iii}] emission are PNe. However, when we consider the spatial scales covered by one PSF FWHM ($\sim$ 80pc), we need to verify again that the population of PN that we discover is examined for obvious contamination sources, which primarily are unresolved \ion{H}{ii} regions and SNRs. Previous studies of PNe through on-off band photometry have used band filters ($\sim$ 30 - 60 \AA{} wide) designed to isolate the emission of [\ion{O}{iii}] lines. This would allow for the potential of contamination sources other than those just mentioned, including high-redshift (z$\sim$3.1) Ly$\alpha$ emitting galaxies and background galaxies emitting [\ion{O}{ii}]\,3727\AA{} (z$\sim$0.34).

However, with IFU data, we can resolve the two components of the [\ion{O}{iii}] emission (4959\AA{} and 5007\AA{}). This advantage is helpful in detecting and separating the PNe from impostors. These are sources with only one emission line that are fitted with a dual-peak model and produce a $\chi^{2}$ greater than if the source were a PNe with both emission lines. We therefore relied on the filtering methods discussed here to exclude objects such as Ly$\alpha$ galaxies or [\ion{O}{ii}] background galaxies before the contamination checks. This process was assessed on sources that were identified as single emission peak sources, and it was found to filter such objects out within the fitting and filtering steps.

To address the two other sources of survey contamination, other diagnostic emission lines such as $\rm H\beta, H\alpha$, [\ion{N}{ii}], and [\ion{S}{ii}] must be considered and compared to the emission of [\ion{O}{iii}]. We followed the method of \citet{Kreckel2017AMUSE} and used the ratio of [\ion{O}{iii}] to H$\alpha$ as a primary identifier between PNe sources and compact unresolved \ion{H}{ii} regions \citep[Eq.~\ref{eq:contamination}, see also][]{Ciardullo2002PlanetaryScales, Herrmann2008PlanetaryDistances, Davis2018TheRemnants}. This comparison stems from the fact that the intensity of [\ion{O}{iii}] in PN sources is greater than H$\alpha$.

To identify an SNR, we relied on the initial works of \citet{Riesgo2006RevisedNebulae} and more recently, \citet{Kreckel2017AMUSE}. One key difference in the emission line analysis of SNR compared to \ion{H}{ii} regions is that the [\ion{S}{ii}]-to-H$\alpha$ ratio is higher in SNR than in compact \ion{H}{ii} regions. SNR have been shown to exhibit similar ratios of [\ion{O}{iii}] to H$\alpha$ as PNe \citep{Davis2018TheRemnants} and therefore require their own classifier for identification purposes. Following previous survey methods, we applied a threshold for the ratio of [\ion{S}{ii}] to H$\alpha$, where a source has to exhibit [\ion{S}{ii}] / H$\alpha \ > 0.3$ to be considered an SNR \citep{Blair2004AnM83}. The limiting factor in this approach is that we first have to detect [\ion{S}{ii}] emission with a signal-to-noise ratio of three. This detection is not always possible, however. In these cases, we evaluated an upper value of the emission line ratio when the lines were above a signal-to-noise ratio of three.

We present the results of our contamination analysis in Fig.~\ref{fig:PNe_impostor167} and Fig.~\ref{fig:PNe_impostor219}. Excluded objects are catalogued and given an appropriate ID type in Tables ~4 and ~5. The top panel of Fig.~\ref{fig:PNe_impostor167} shows the flux ratio of [\ion{O}{iii}] and $\rm H\alpha$+[\ion{N}{ii}], plotted against $\rm m_{5007}$, along with the limits set out in Eq.~\ref{eq:contamination} \citep{Ciardullo2002PlanetaryScales},
\begin{equation}
    4 < \log_{10} \bigg( \frac{F([\ion{O}{iii}])}{F(H\alpha + [\ion{N}{ii}])} \bigg) < -0.37 M_{5007} - 1.16.
\label{eq:contamination}
\end{equation}
Here, we find a number of sources below the cone region, highlighting sources with a higher-than-expected abundance of H$\alpha$+[\ion{N}{II}] in comparison to [\ion{O}{iii}] for a given $m_{5007}$. The data points are colour-coordinated with respect to the PN $\rm A_{H\alpha}/rN$ level. Sources with detected [\ion{S}{ii}] emission with a signal-to-noise ratio higher than 2.5 of the residual noise are highlighted by a circle. Only one [\ion{O}{iii}] emitting source, detected within FCC\,167, is found to emit [\ion{S}{ii}] above this threshold. The lower panel of Fig.~\ref{fig:PNe_impostor167} presents the second impostor check. This panel has a few juxtaposed regions that help identify where certain sources would appear based on the ratios of various emission lines. For FCC\,167, we identified five objects that were excluded: four highly likely potential SNRs, and one object that probably is a compact \ion{H}{ii} region.
We ran the same impostor checks on FCC\,219, with the results displayed in Fig.~\ref{fig:PNe_impostor219}, but found no sources with an [\ion{S}{ii}] signal above 2.5 times the background. We find seven objects that were excluded: three highly likely SNRs, and four likely \ion{H}{ii} regions objects. We also note that the lower panel of Fig.~\ref{fig:PNe_impostor219} contains objects within the \ion{H}{ii} area that are not below the defining limit of the upper panel. These are probably PN because PNe tend to overlap in a region where several PNe have been observed with greater H$\alpha$ emissions than the main population of PNe.

We are confident that we reliably excluded impostor sources without confusing background diffuse emission with that originating from the unresolved point source. This is aided by imposing the same line profile width, as fitted from [\ion{O}{iii}], to the other fitted emission lines. This certifies that we do not report line strengths from background diffuse ionised gas. The line profiles of background emissions like this would appear wider than those originating from unresolved point sources moving with the stars.
Fig.~\ref{fig:gandalf_SNR} displays our GandALF fit for the brightest of our SNR sources. There, we highlight a few regions of particular interest, which are the regions around the [\ion{O}{iii}] 4956 5007\AA \AA{} (bottom left), H$\alpha$ 6563\AA{}, and [\ion{N}{ii}] 6548 6583\AA{}\AA{} [\ion{S}{ii}] 6716 6731\AA{}\AA{} (bottom right) nebular emission lines.


\section{Results and discussion}
\subsection{Results for PNe}
Within the central region of FCC\,167, we catalogue 91 [\ion{O}{iii}] emitting sources, labelled PNe. Table 4 summarises the outcome from our filtering procedure. It presents our catalogue of the PNe sources, which are also plotted in Fig. \ref{fig:FCC167_A_rN}, highlighted with black circles, with the over-luminous source that is highlighted by a black square icon. The PNe we found to match those reported in \citet{Feldmeier2007CalibratingResults} are highlighted by a blue square. All PNe were labelled with their identifying number. In addition, the table contains their RA and DEC (J2000), apparent magnitude in [\ion{O}{iii}] ($\rm m_{5007}$), and $A/rN$.

Of the detected [\ion{O}{iii}] emitting sources within the FOV of FCC\,167, one source appears over-luminous by 0.4 mag with respect to the predicted cut-off of the PNLF. This is expected, and sources like this have previously been reported \citep{Jacoby1996PlanetaryRegion, Longobardi2013TheM87}. A few scenarios have been put forward to account for over-luminous sources. One possibility would be a chance superposition of a number of PNe. Another statistically more favoured possibility is that such objects are the product of coalesced binaries. Our emission line filtering did not allow for the safe identification of this particular source, but it indicated that such sources are less likely to be due to an \ion{H}{ii} region or an SNR. This observation is further supplemented by the fact that FCC\,167 is a typical early-type galaxy with an older stellar population, which is expected to be dominated by low-mass stars ($\sim 1 M_{\sun}$). Within such a population we do not expect to have either very luminous \ion{H}{ii} regions or frequent SN explosions. Moreover, our spatial resolution spans $\sim$80pc, and blending of PNe sources is therefore quite likely.

In the central observation of FCC\,219, we catalogue 56 [\ion{O}{iii}] emitting sources and classify them as PNe. Of the originally detected point sources, five were deemed to be PNe impostors, although no interlopers are present. We note that the three sources classified as SNR are closely grouped together, which may mean that the underlying stellar environment was part of a more recent star formation burst, and that the NS may be type II. Figure~\ref{fig:FCC219_A_rN} shows the FOV, and the catalogued sources are highlighted with black circles, and those that matched sources of \citet{McMillan1993PlanetaryCluster} are again highlighted by a blue square. See Table~5 for the properties of the catalogued sources.

\subsection{PNLF}
The empirical form of the PNLF \citep[as introduced by][Eq.~\ref{eq:PNLF}]{Ciardullo1989PlanetaryCompanions} is described by an exponential drop-off at the bright end, with an accompanying exponential tail for the fainter end. It can be approximated with the following functional form:
\begin{equation}
    N(M) \propto e^{ \ c_1 \ M_{5007}} \ \big[1 - e^{ \ 3( \ M^{*}_{5007} - \ M_{5007} \ )}\big],
    \label{eq:PNLF}
\end{equation}
where $\rm M_{5007}$ is the absolute magnitude of the detected PN. $M^{*}_{5007}$ is the cut-off absolute magnitude of the brightest PNe, originally calibrated to $-$4.48, from observations of the M31 PNe population, assuming a Cepheid distance of 710kpc \citep{Ciardullo1989PlanetaryCompanions}. The $c_1$ parameter details the behaviour of the function tail and was derived from the model of an expanding ionised [\ion{O}{iii}] spherical shell \citep[$c_1$ = 0.307;][]{Henize1963DimensionsCloud.}.
More recently, \citet{Ciardullo2012TheGaia} explored the PNLF zero-point ($M^{*}_{5007}$), calibrating its value with galaxies that already had distance estimates from Cepheids and/or the method called  tip of the red giant branch (TRGB). He reported an agreement between these two distance estimators with $M^{*}_{5007}=-4.53$, but also failed to find evidence for a metallicity dependence on $M^{*}_{5007}$. \citet{Gesicki2018TheFunction} further explored the effect of star formation history and stellar population age on the bright cut-off of the PNLF of a detected PNe population. Together with the recent work of \citet{Valenzuela2019RevisedFunction}, these studies finally overcome the initial difficulties of obtaining $M^{*}_{5007}$ PNe in old stellar populations with mostly solar-mass star progenitors \citep[e.g.][]{Marigo2004EvolutionPNLF}.

Another area of particular interest is the faint end of the PNLF. Observations of Local Group galaxies, including the Large and Small Magellanic Clouds, have so far yielded the exploration of the faint end of the PNLF because such surveys would cover a greater magnitude range than galaxies beyond $\sim$10 Mpc. Surveys such as ours are limited to the bright end of the distribution, exploring $\sim 1-2$ mag down from the cut-off. It has been demonstrated within M31 by exploring $\sim 5-6$ mags from the cut-off that the fainter end does increase in number \citep{Bhattacharya2019TheCFHT}. The authors commented that this might be attributed to an older stellar population, while the bright end is dominated by a younger stellar population, formed 2-4 Gyr ago. For our investigation it is therefore imperative to understand and account for incompleteness, which underlines our conclusions about the observed PNLF.

\placefigPNLFa
\placefigPNLFb

To construct our PNLF, we resampled our PNe into 0.2 mag bins and estimated a distance modulus by assuming that the brightest source of our sample is located at the bright cut-off, $M_{5007}^*=-$4.53 (taking the second brightest object in FCC167, see below). With this distance modulus, we shifted the PNLF form predicted by \citet{Ciardullo1989PlanetaryCompanions}  so that it can be directly compared to our observed PNLF. This is clearly an incomplete comparison because fainter objects are expected to be lost to the noise of our spectra from either the sky background or the stellar continuum. Furthermore, closer to the centre of the galaxy, even the brightest PNe may become undetectable, and ionised gas might occasionally complicate things further.

Within the region we considered to construct our PNLF (i.e. excluding masked regions), we defined the detection completeness at any given $m_{5007}$ magnitude as the fraction of stellar light contained in the area where a PNe of that magnitude can be detected, similar to \citet{Sarzi2011TheView} and \citet{Pastorello2013TheView}. According to our PNe detection criteria, this area includes the MUSE spaxels where the PNe peak spectral $A_{[\ion{O}{iii}]}$ amplitude for a PNe of $m_{5007}$ magnitude would exceed three times the local spectral residual noise level (rN) from our residual datacubes.

After computing the completeness level as a function of $m_{5007}$ , we used this function to produce an incompleteness-corrected \citet{Ciardullo1989PlanetaryCompanions} form of the PNLF. This was then rescaled so that its integral matched the total number of PNe in the observed PNLF.

This correction process draws on the assumption that the observed PNe are indeed drawn from the \citet{Ciardullo1989PlanetaryCompanions} form of the PNLF. That this incompleteness-corrected model PNLF matches the observed PNLF of FCC167 and FCC219, shown in Fig.~\ref{fig:FCC167_PNLF} and \ref{fig:FCC219_PNLF}, respectively, well appears to already validate our assumption. A simple Kolmogorov-Smirnov test yields a p-value = 0.99 for FCC\,167 and FCC\,219, indicating in both cases that we cannot reject the null hypothesis that the observed PNLF is drawn from the incompleteness-corrected \citet{Ciardullo1989PlanetaryCompanions} form.

\subsection{Distance estimation}
From the invariant cut-off in PNe luminosity that is observed in different galaxies and galaxy types, we can derive a distance by converting the apparent magnitude into absolute magnitudes through the distance modulus. This distance estimation compliments the wide set of other methods such as the surface brightness fluctuation (SBF), SN type Ia (SNIa), and TRGB and fundamental plane (FP). To use the detected PNe as a distance estimator, a sufficiently large sample is required that has to contain the brightest PNe.

For FCC\,167, returning to the over-luminous object, when we assume that the source is at the cut-off for our PNe population ($M^{*}_{5007}$), the resulting distance modulus estimate is 30.9 (15.2\,Mpc). This is at odds with our PNLF distance and with previous distance estimates of the Fornax cluster as a whole. The cluster distance is estimated to lie between 17--22\,Mpc \citep{Ferrarese2000TheConstant, Blakeslee2009THEDISTANCE, Tully2016COSMICFLOWS-3}. Another discrepancy introduced by including this overly bright source is an apparent lack of intermediately bright PNe. The over-luminous source $m_{5007}$ is distinctly 0.4 mag brighter than the rest of the population, and no intermediate PN are detected. For these reasons, we decided to omit this source from our catalogue and further analysis.

Assuming that the brightest PN of our filtered sample resides at the bright cut-off for the PNLF, we find a distance modulus of 31.24$\pm 0.11$ ($D_{PNLF} = 17.68 \ \pm \ 0.91$ Mpc). This value agrees with \citet{Tully2016COSMICFLOWS-3} and \citet{Ferrarese2000TheConstant}, who reported values of 31.35 $\pm$ 0.24 from SNIa, SBF, and FP measurements, and 31.37 $\pm$ 0.20 from SBF and globular cluster LF, respectively. If, on the other hand, we were to evaluate the 91 PNe candidates at a pre-determined distance of 21.2 Mpc (31.63 $\pm$ 0.08) (\citep{Blakeslee2009THEDISTANCE}, who determined their distance from SBF measurements in the Sloan z-band), the distribution of PNe shifts towards the more luminous end of the PNLF, reaching $M_{5007} \approx -5.0$. This would contradict most if not all previous PN surveys, which found $M^{*}_{5007} \approx -4.5$.

For the central observation of FCC\,219, we estimate a distance modulus of 31.42 $\pm$ 0.1 (19.24 $\pm$ 0.84 Mpc). This agrees within the stated limits with the distance from \citet{Tully2016COSMICFLOWS-3}, 31.37 $\pm$ 0.22, and also with the measurements reported by \citet{Ferrarese2000TheConstant}, 31.22 $\pm$ 0.12. We conclude that our results lie within the distance range expected for this cluster, and note that previous studies have found some discrepancies between SBF and PNLF estimates \citep{Ciardullo2012TheGaia, Kreckel2017AMUSE}.

\subsection{Luminosity-specific PN frequency}

As discussed in Sect.~5.2 and shown in Figs. 16 and 17, the observed PNLFs in FCC167 and FCC219 are consistent with the \citet{Ciardullo1989PlanetaryCompanions} empirical form of the PNLF when this is corrected for incompleteness and rescaled so that its integral matches the total number of PNe we observed. By applying the same normalisation also to the original function, we can similarly integrate it to estimate the total number of PNe in the regions under consideration, $N_{PNLF,\, \Delta M}$ , down to some magnitude limit $\Delta M$. Dividing this number by the stellar bolometric luminosity in the same region ($L_{Bol}$), we can arrive at the luminosity-specific PN frequency,
\begin{equation}
    \alpha_{\Delta m} = N_{PNLF,\,\Delta m} / L_{Bol},
    \label{eq:alpha}
\end{equation}
which may depend on the stellar parent population properties \citep[e.g.][]{Buzzoni2006PlanetaryPopulations}. In this paper, we rely on the commonly adopted $\alpha_{2.5}$ measurement for the luminosity-specific PN frequency, where the PNLF is integrated down to 2.5 magnitudes from the bright-end cut-off point.

To evaluate the bolometric luminosity, we first proceeded to obtain an integrated spectrum of the stellar regions where the presence of PNe was investigated (e.g. within the MUSE FOV, excluding the masked region). We then fit this spectrum with pPXF using the EMILES templates \citep{Vazdekis2016UV-extendedGalaxies}, using the resulting template weights to reassemble an optimal template for the entire wavelength range of the EMILES templates. Because this is rather extended, we can compare the total flux to that in the SDSS g-band to determine a bolometric correction for the g-band. We can then apply this to the g-band flux observed in our integrated spectra and thence obtain a bolometric luminosity at the distance derived from the PNLF. Table~\ref{table:alpha} presents $N_{PNLF}$ and $L_{bol}$ that we used to determine $\alpha_{2.5}$.

Our estimates of $\alpha_{2.5}$ are somewhat different for the two galaxies, where the value of FCC\,167 is $\sim$1.6 times that of FCC\,219. According to \citet{Buzzoni2006PlanetaryPopulations}, this difference in the specific number of PNe may be related to a difference in metallicity in the parent stellar population. However, in the case of FCC\,167 and FCC\,219, \citet{Iodice2019TheMaps} reported rather similar values for the central (within an effective radii, 0.5 $R_{e}$) stellar metallicity ([M/H]): 0.09 dex for FCC\,167 and 0.14 dex for FCC\,219. One potential source of this difference might be the hot gas halo in FCC\,219, as demonstrated by X-ray data \citep[e.g.][]{Jones1997XRayCluster, Murakami2011SuzakuDistribution}. However, FCC\,167 does not posses as significant a hot-gas inter-stellar or  inter-galactic medium component. The lower $N_{PNLF}$ value reported for FCC\,219 might therefore stem from the ram pressure the PNe gas would experience as it passes through the hot medium. This would naturally act to sweep the PNe gas associated with a central ionising star \citep{Conroy2015PreventingHeating}. \citep{}. More effort from a modelling perspective is still needed to elucidate the potential effect that hot X-ray halos might have on the population of PNe \citep[e.g. following ][]{Li2019TheICM}. Earlier studies of this interaction focused on the Virgo cluster, M87, and found evidence of re-compression of PN shells for PNe closer to the galactic nucleus \citep{Dopita2000DoNebulae}. For the PN that are ejected into the intra-cluster medium, shorter evolutionary times ($\tau_{PN}$) than previous estimates \citep{Villaver2005TheMedium} were reported.

\begin{table}
\renewcommand{\arraystretch}{1.5}
\caption{Galaxy, number of PNe according to PNLF, total bolometric luminosity, and $\alpha_{2.5}$.}
\label{table:alpha}
\begin{tabular}{l l l l}
\hline
Galaxy & $N_{PNLF,\,2.5}$ & $L_{Bol}$ & $\alpha_{2.5}$ \\
&  & ($10^{9} \ L_{\sun}$) & ($10^{-8} \ N_{PNe} \ L_{\sun}^{-1}$) \\
\hline
FCC 167 & 277 $\pm$ 29 & 16.99$^{+1.82}_{-1.51}$ & $1.63^{+0.24}_{-0.23}$ \\
FCC 219 & 287 $\pm$ 38 & 27.13$^{+2.49}_{-2.21}$ & $1.06^{+0.16}_{-0.17}$ \\
\hline
\end{tabular}

\end{table}

\section{Conclusions}

We have attempted to achieve a consensus in detecting PNe on the example of the population of PNe in two galaxies (FCC\,167 and FCC\,219). For this purpose, we developed a novel fitting and detection procedure that is capable of combining the spectra and spatial information contained in our IFU observations. In this, the spatial information is portrayed by a Moffat distribution function. We further demonstrated that it is capable of also successfully accounting for the [\ion{O}{iii}] spectral emission lines. With our procedure we were either able to fix or to infer the instrumental PSF. Moreover, we ran an extensive set of simulations to constrain its limitation, in particular, concerning our MUSE observations. We illustrated the capabilities of this newly developed procedure by applying it to two galaxies.

The primary outcomes of the analysis carried out on FCC\,167 and FCC\,219 include the points listed below.

\begin{itemize}
    \item We made catalogue of 91 detected [\ion{O}{iii}] unresolved point sources, characterised here as PNe in origin, within FCC\,167 and a catalogue of 56 PNe within FCC 219.

    \item Through our modelling techniques, we accurately reproduced the PSF of each pointing after fitting multiple PNe in parallel with the same PSF shape. This improves the accuracy of the reported [\ion{O}{iii}] flux values.

    \item Through simulations of modelling known PNe, we tested and verified the reliability of our results and the accuracy of the parameters we used within the 1D+2D modelling technique. This investigation also highlighted the limitations in A/rN that must be factored in when filtering for outliers. We are confident in the results of our method for categorising sources as PNe when measured above A/rN of 3.

    \item Through emission line ratio diagnostics and by comparing the velocity of the PNe to the background stellar populations, we identified one interloper and five potential impostors within our FCC\,167 sample: four SNR, and one compact \ion{H}{ii} region. For FCC\,219, we identified three SNR and four likely compact \ion{H}{ii} regions, with no evidence of any interloping PNe within the catalogue.

    \item We calculated the values for the luminosity-specific PN frequency, $\alpha_{2.5}$, for the population of PNe down to 2.5 mag from the bright-end cut-off for FCC 167 and FCC 219: $1.63^{+0.24}_{-0.23} \times 10^{-8}$ and $1.06^{+0.16}_{-0.17} \times 10^{-8}$ , respectively.

    \item Finally, through the use of the PNLF and the invariant cut-off in brightness, we report distances to the host galaxies for the PNe: 17.68 $\pm$ 0.91 Mpc for FCC\,167, and 19.24 $\pm$ 0.84 Mpc for FCC\,219. The two distance estimates agree with the literature and are consistent with other methods that used SBFs and SNIa measurements. They also agree within the limits with the distance to the Fornax cluster (17--22Mpc)
\end{itemize}

Moving forward, we wish to explore the remaining bright early-type galaxy population within the Fornax 3D survey. We will catalogue the positions, magnitudes, and emission line ratios of their PNe populations within the central regions. When the catalogue of early-type galaxies has been evaluated, we will compare distance estimates from the PNLF with other current methods. The primary scientific analysis will consist of comparing the $\alpha_{2.5}$ value of each galaxy with their relative galactic properties: UV excess, metallicity, and other such properties that might affect stellar evolution and hence PNe formation.

\begin{acknowledgements}
We would like to thank the referee for their constructive responses, which has improved the content and clarity of this manuscript. Based on observations collected at the European Southern Observatory under ESO programme 296.B-5054(A). TS thanks A. Bittner, A. Jones and R. Jackson and M. Timberhill for their help and ideas that aided in the construction and presentation of this paper. This work was supported by Science and Technology Facilities Council [grant number ST/R504786/1]. SV acknowledges the support of the Flemisch Research Fund (FWO). RMcD is the recipient of an Australian Research Council Future Fellowship (project number FT150100333). J. F-B acknowledges support through the RAVET project by the grant AYA2016-77237-C3-1-P from the Spanish Ministry of Science, Innovation and Universities (MCIU) and through the IAC project TRACES which is partially supported through the state budget and the regional budget of the Consejer\'\i a de Econom\'\i a, Industria, Comercio y Conocimiento of the Canary Islands Autonomous Community. GvdV acknowledges funding from the European Research Council (ERC) under the European Union's Horizon 2020 research and innovation programme under grant agreement No 724857 (Consolidator Grant ArcheoDyn). This project made use of the following software packages: LMfit \citep{Newville2014LMFIT:Python, Newville2019Lmfit/lmfit-py1.0.0}, Astropy, a community-developed core Python package for Astronomy \citep{Robitaille2013Astropy:Astronomy}, scipy \citep{EricJonesandTravisOliphantandPearuPetersonandetal.2001SciPy:Python}, Numpy \citep{VanDerWalt2011TheComputation}, matplotlib \citep{Hunter2007Matplotlib:Environment} and Pandas \citep{Mckinney2010DataPython}.
\end{acknowledgements}


\bibliographystyle{aa}
\bibliography{references}

\end{document}